\documentclass[fleqn,usenatbib]{mnras}

\usepackage{newtxtext,newtxmath}

\usepackage[T1]{fontenc}
\usepackage{float}
\usepackage{physics}
\usepackage{subfiles}
\usepackage{comment}
\usepackage{booktabs}
\usepackage{graphics}
\usepackage{comment}
\usepackage{graphicx}
\usepackage{hyperref}
\usepackage{xcolor}
\usepackage{lipsum} 
\usepackage{tabularx}
\usepackage{graphicx}
\usepackage{orcidlink}
\DeclareRobustCommand{\VAN}[3]{#2}
\let\VANthebibliography\thebibliography
\def\thebibliography{\DeclareRobustCommand{\VAN}[3]{##3}\VANthebibliography}

\usepackage{graphicx}	
\usepackage{amsmath}	
\usepackage{xcolor}

\definecolor{blue_cust}{HTML}{0066ff}

\definecolor{gold_cust}{HTML}{FFBF00}

\definecolor{purple_cust}{HTML}{800080}

\definecolor{pink_cust}{HTML}{DC267F}

\definecolor{perfect_green}{HTML}{4FBF26}

\definecolor{crimson}{HTML}{DC143C}

   \title[Radiative and Collisional Atomic Data for Sr {\sc ii} and Y {\sc ii}]{New Radiative and Collisional Atomic Data for Sr {\sc ii} and Y {\sc ii} with application to Kilonova modelling}
   
\author[L.P. Mulholland et al.]{
L. P. Mulholland,\orcidlink{0009-0003-2668-5589}$^{1}$\thanks{E-mail: lmulholland25@qub.ac.uk}
N. E. McElroy,\orcidlink{0009-0002-7315-5444} $^{1}$
F. L. McNeill,\orcidlink{0009-0001-9528-7475}$^{1}$
S. A. Sim,\orcidlink{0000-0002-9774-1192}$^{1}$
C. P. Ballance \orcidlink{0000-0003-1693-1793}$^{1}$
\newauthor
\hspace{1mm}and
C. A. Ramsbottom \orcidlink{0000-0003-1579-8556}$^{1}$
\\
$^{1}$Astrophysics Research Centre, School of Mathematics \& Physics, Queens University Belfast, BT7 1NN, Northern Ireland.
}

\date{Accepted XXX. Received YYY; in original form ZZZ}

\pubyear{\the\year{}}

\begin{document}
\label{firstpage}
\pagerange{\pageref{firstpage}--\pageref{lastpage}}
\maketitle

\begin{abstract}
   {The spectra of singly ionised Strontium and Yttrium (Sr {\sc ii} and Y {\sc ii}) have been proposed as identifications of certain spectral features in the AT2017gfo spectrum. With the growing demand for NLTE simulations of Kilonovae, there is a increasing need for atomic data for these and other $r$-process elements.}
   {Our goal is to expand upon the current set of  atomic data for $r$-process elements, by presenting transition probabilities and Maxwellian-averaged effective collision strengths for Sr {\sc ii} and Y {\sc ii}. }
   {The Breit-Pauli and DARC $R$-matrix codes are employed to calculate the appropriate collision strengths, which are thermally averaged according to a Maxwellian distribution to calculate excitation and de-excitation rates. The {\sc tardis} and {\sc ColRadPy} packages are subsequently used to perform LTE and NLTE modelling respectively.}
   {A complete set of transition probabilities and effective collision strengths involving levels for Sr {\sc ii} and Y {\sc ii} have been calculated for temperature ranges compatible with kilonova plasma conditions. Forbidden transitions were found to disagree heavily with the Axelrod approximation, an approximation which is currently employed by other models within the literature. Theoretically important spectral lines are identified with both LTE and NLTE modelling codes. LTE simulations in {\sc tardis} reveal no new significant changes to the full synthetic spectra. NLTE simulations in {\sc ColRadPy}  provide indications of which features are expected to be strong for a range of regimes, and we include luminosity estimates. Synthetic emission spectra over KNe densities and temperatures reveal potentially interesting spectral lines in the NIR.}
\end{abstract}




\section{Introduction}

The merger of neutron stars is speculated to be a major source of rapid neutron-capture (r-process) elements. This has been proposed theoretically in publications such as \cite{Latt77} and confirmed observationally by a series of X-Shooter spectra for the AT2017gfo kilonova (KNe) taken daily from about 1.5 days post merger (\cite{Smartt2017}; \cite{Pian17}). These series of spectra cover wavelengths from the ultraviolet to the near-infrared and track the evolution of absorption and emission features in the transient. Confirmation of a well identified P Cygni spectral line of Sr {\sc ii} at approximately 1$\mu$m was found by \cite{Watson19} and a further discovery of a 760nm Y {\sc ii} P Cygni line in AT2017gfo was identified by \cite{SnepWat23}. These findings are important as P Cygni lines provide vital information on the composition of the merger ejecta, the velocity of the explosion and also act as a probe to investigate the geometry and abundance stratification of the KNe ejecta. These two recent publications provide further motivation for the work presented here. 

To accurately model the neutron star merger (NSM) in either Local Thermodynamic Equilibrium (LTE) or non-LTE (NLTE), a significant amount of atomic data pertaining to the ion stages of interest are necessary inputs for the radiative transfer codes. Inputs such as energy levels and Einstein A-coefficients are required for basic LTE analysis with the more computationally demanding collisional data for excitation, ionization and recombination required for NLTE radiative transfer modelling. It is now well known that as the NSM ejecta expands with increasing time and decreasing density and temperature, the KNe transitions from Saha equilibrium LTE to NLTE and in order to investigate its spectral evolution a full NLTE treatment is required. \cite{Pognan2023} showed that even as little as a few days post merger the disparities between LTE and NLTE opacities could be as large as several orders of magnitude for specific r-process ion stages. In addition, NLTE modelling of Au {\sc i} by \cite{McCann2022} identified that the metastable level populations can deviate quite substantially from their LTE proportions at conditions relevant to NSM events. Indeed excitation from these metastable levels can dominate direct excitation from the ground state. These changes coincide with the transition from an absorption to an emission spectrum marking the changeover to the nebular phase (\cite{Hotokezaka2021}) where the conditions are also insufficient to maintain LTE level populations.

A major hindrance to the KNe modelling is the lack of complete and spectroscopic datasets for the r-process ions of interest, particularly in the availability of electron-impact excitation rates, photoionization cross sections and radiative recombination rates. For many of the high-Z species of interest there is either a significant paucity of data available or no data at all. There has, however, been a flurry of activity in the community since the recent AT2017gfo KNe event in an attempt to rectify this. A good example is the systematic opacity calculations of \cite{Tanaka20} who computed atomic structure calculations for the first three ionization stages of the r-process elements from Fe (Z=26) to Ra (Z=88). In this work the {\sc hullac} code was utilised to calculate the energy levels and transition rates used to compute the bound-bound opacities of r-process elements to ultimately understand the elemental variation in neutron star merger events. The authors, however, do not claim to provide spectroscopic accuracy for the atomic structure data, their goal was to provide a complete set of data for opacity calculations. The detailed spectral features in the X-Shooter spectra of AT2017gfo can only be identified correctly with accurate, spectroscopic atomic datasets. This point was emphasised by \cite{Shingles2023} who presented self-consistent 3D radiative transfer simulations of NSM spectra. They found that major differences appeared in the resulting KNe synthetic spectra when spectroscopic energies were adopted in the models for ions in the first r-process peak, Sr, Y and Zr.

For the two ions of interest in this paper, Sr {\sc ii} and Y {\sc ii}, a full literature search reveals the following available data. In the National Institute of Standards and Technology (NIST) ASD database (\cite{nist}) there are 71 energy levels and 33 transition probabilities for lines between 178.4 - 1091.5nm for Sr {\sc ii}. The energy levels and A-values were compiled from the works of \cite{San12}, \cite{Moore52} and \cite{WieMar80}. A re-evaluation of some transition probabilities was subsequently performed by Kramida (unpublished 2016) for ASD v5.4. Collisional data for Sr {\sc ii} is much more scarce. An early LS calculation by \cite{Burgess1989} produced Maxwellian averaged rate coefficients from distorted-wave collision strengths for transitions among the lowest three LS terms only. \cite{Duan2013} reported electron-impact broadening parameters and shifts for some spectral lines in Be {\sc ii}, Sr {\sc ii} and Ba {\sc ii} required for elemental abundance analysis in solar and stellar atmospheres. The most useful atomic dataset currently available in the literature was produced by \cite{Bautista2002} to investigate Sr {\sc ii} emission in the ejecta of Eta Carinae. Excitation rates and radiative transition probabilities were computed for transitions among the lowest 12 levels of Sr {\sc ii}. Unfortunately the datasets for both the collision strengths and effective collision strengths were not published and the temperature range for the excitation rates was 5000 - 20 000K, higher than those required for the current KNe modelling.

For Y {\sc ii} considerably more radiative data is available in NIST, 236 energy levels experimentally measured by \cite{Nilsson91} and A-values for 66 lines with wavelengths between 311.2 and 788.2nm. The transition probabilities were measured by \cite{Hannaford82} by combining laboratory measurements of radiative lifetimes and branching ratios. Interestingly, the most prominent Y {\sc ii} line identified by \cite{SnepWat23} is located at 788.19nm just on the edge of the available NIST data. In addition, \cite{Biemont11} used the Hartree-Fock method including core-polarization effects (HFR+CPOL) to compute oscillator strengths and A-values for 84 transitions in Y {\sc ii} in the wavelength range between 224.3 and 1060.5nm. These calculations were extended using the same method by \cite{Palmeri17} to include configuration interaction terms up to n=10 to compute radiative data in the extended wavelength range 194 to 3995nm. It should be noted that experimental lifetimes of the upper levels were used in this work to rescale the A-values and it is these rescaled values that are recommended by the authors. Collisional data for excitation and ionization is scarce for Y II. \cite{Smirnov2001} measured excitation cross sections for 32 spectral lines of Y II at energies up to 200eV. The spectral range considered in this work was 224-540nm. \cite{Pognan2023} performed NLTE KNe modelling using excitation rates computed using the  \cite{Axelrod1980} and  \cite{VanReg1962} approximations. The use of such approximations has been subsequently shown by \cite{Bromley2023}, for all three ionization stages of Pt, to systematically underestimate the effective collision strengths for both forbidden and allowed lines. The paper concludes that such approximations are not well suited for use in KNe-like modelling conditions, at least for Pt {\sc i}, {\sc ii} and {\sc iii}. In addition, collisional radiative models that incorporated these Axelrod and van Regemorter rates systematically underpopulated the metastable levels and overpopulated the excited states when compared to data that was produced by a full close-coupling $R$-matrix treatment. These findings again provide compelling evidence and a strong motivation for the current work.

It should also be emphasised that the atomic data presented in this work has applications outside of the NSM modelling and will be useful in several diverse research areas. For example, a recent publication by \cite{StormBerg2023} performed a NLTE analysis of strong diagnostic lines of Y {\sc ii} in the spectra of cool stars and discuss the implications of Y as a galactic chemical clock. A further paper by \cite{Storm2024} measured solar abundances of Y and Eu and found that including 3D NLTE effects in the Y II lines increased the solar Y abundance by $\sim$ 0.2 dex compared to 1D LTE. In addition, \cite{Alex2023} determined NLTE Y abundances in a large sample of 65 well studied F-G-K dwarfs and subgiants to use in chemical evolution studies and heavy element enrichment of the interstellar medium. Finally, \cite{Naslim2011} detected lines from the low ionization stages of the r-process elements Sr, Y, Zr and Ge in the spectrum of the hot subdwarf LSIV-14$^{\rm{o}}$116 which yielded measured abundances between 3 and 4.6 dex above the solar value. These overabundances have since been confirmed for other hot subdwarfs Feige46 and PHL417 by \cite{Dorsch2020} which raises the question of the contribution from these elements to the opacity and structure of the stellar atmosphere. There has been additional effort to spectroscopically study metal-poor stars with $r$-process signatures in dwarf galaxies. Notable examples include the faint dwarf galaxy Reticulum II \citep{FREBEL2019167909} and with strong Sr absorption observed in J0246-1518 \citep{hansen2018ages,Frebel2023}.

The present publication is structured as follows. In Section 2 the atomic models used in the Sr {\sc ii} and Y {\sc ii} computations are discussed. Radiative data such as target energy levels and A-values for transitions among these levels are compared with all known values currently available in the literature. In Section 3 collisional cross sections from the electron-impact excitation scattering calculations are presented for a selection of transitions as well as Maxwellian averaged effective collision strengths for temperatures of relevance to KNe modelling. A comprehensive dataset of every Maxwellian averaged transition shall be made availiable at \cite{openadas_site}.   In Section 4 new 1D \textsc{tardis} LTE radiative transfer spectra computed using this new atomic data, are presented and discussed in detail. In Section 5 the atomic data for both systems are included in a full NLTE collisional radiative model to investigate the population dynamics of the levels, potential temperature and density diagnostic lines as well as the photon emmissivity co-efficients for a wide range of wavelengths. Finally in Section 6 a summary and some conclusions are drawn. 

\section{Atomic Structure - Sr {\sc ii} and Y {\sc ii} Models}
The target models used to generate the Sr {\sc ii} and Y {\sc ii} energy levels and Einstein A-values for transitions among these levels were computed in this work using either the fully relativistic {\sc grasp}$^0$ or the semi-relativistic {\sc autostructure} packages. Several models were investigated for both systems using both codes and the best model for each was retained for the collision computations. 

The {\sc grasp}$^0$ package was developed initially by \cite{Grant80} and later published by \cite{Dyall1989}. This fully relativistic code solves the multi-configurational Hartree-Fock equations to determine an optimised set of atomic orbitals using the variational principle to compute the minimum energy of the Hamiltonian. By
implementing the Multi-Configuration Dirac Hartree Fock Method (MCDHF) we can solve the Time Independent Dirac Equation (TIDE),
\begin{equation}
     H_{D}\phi=E\phi,
\end{equation}
to obtain the Dirac wavefunctions and the energy eigenvalues $E$ of the Hamiltonian, $H_D$ given by (in atomic units)
\begin{equation}
       H_{D}=\sum_{i=1}^{N} \left(c\boldsymbol{\alpha}\cdot \boldsymbol{p}_i+(\beta-\mathbf{I}_4)c^2-\frac{Z}{r_i}\right)+\sum_{i>j=1}^{N}\frac{1}{r_{ij}}.
    \label{dirac3}
\end{equation}
 where $\boldsymbol{\alpha}$ and $\beta$ are related to the set of Pauli spin matrices, $\boldsymbol{I}_{4}$ is the 4 $\times$ 4 identity matrix, $Z$ is the atomic number, $c$ is the speed of light, $\boldsymbol{p}$ is the momentum operator defined as $\boldsymbol{p}=-i\nabla$, $r_i$ denotes the position of electron $i$ and $r_{ij}$ = $\mid r_i  - r_j \mid$ is the inter-electronic distance. In this work, the Extended-Average-Level method is employed  which weights the diagonal elements according to the statistical weight $2J+1$. A variational procedure optimises the trace of the Hamiltonian. 

The semi-relativistic {\sc autostructure} package was initially developed by \cite{EISSNER1991} but was significantly modified by \cite{Badnell86} and \cite{Badnell97}. This code incorporates an N-electron Breit-Pauli Hamiltonian given by
\begin{equation}
H_{BP}=H_{NR}+H_{RC}
\end{equation}
where $H_{NR}$ contains the non-relativistic operators
\begin{equation}
     H_{NR}   = \sum \limits_{i=1}^{N}\begin{pmatrix}  -\frac{1}{2}\nabla_{i}^{2} - \frac{Z}{r_i} \end{pmatrix} + \sum \limits_{i>j=1}^{N}\frac{1}{r_{ij}},
\end{equation}
and $H_{RC}$ contains the one-body relativistic correction operators (mass correction, Darwin and spin-orbit) given by
\begin{equation}
   H_{RC}= \frac{\alpha^2Z}{2} \sum \limits_{i=1}^{N} \frac{\boldsymbol{l_i} \cdot \boldsymbol{s_i}}{r_{i}^3}  -\frac{\alpha^2}{8} \sum \limits_{i=1}^{N}\nabla_{i}^4  -\frac{\alpha^2Z}{8} \sum \limits_{i=1}^{N}\nabla_{i}^2\begin{pmatrix}\frac{1}{r_i}\end{pmatrix},
\end{equation}
where $\boldsymbol{l_i}$ and $\boldsymbol{s_i}$ are the single-electron orbital and spin angular momentum operators respectively. Adopting a Thomas-Fermi-Dirac-Amaldi model potential with scaling parameters 
$\lambda_{nl}$ for each $nl$ orbital the {\sc autostructure} code generates a set of orbital parameters for use in the structure calculations. 

\begin{table}
    \centering
    \begin{tabular}{c c l l }
    \hline \\
    \multicolumn{4}{c}{ Structure Models } \\
    \hline \\
    Sr {\sc ii} Model  && 4p$^{6}$4d & 4p$^{6}$4f \\
     - 22 configs      && 4p$^{6}$5s & 4p$^{6}$5p \\
       {\sc autostructure}      && 4p$^{6}$5d & 4p$^{6}$6s  \\
                       && 4p$^{6}$6p & 4p$^{6}$6d \\
                       && 4p$^{6}$7s & 4p$^{6}$7p \\
                       && 4p$^{6}$7d & 4p$^{6}$5f \\
                       && 4p$^{6}$5g & 4p$^{6}$8s \\
                       && 4p$^{5}$5s4d   & 4p$^{5}$5s4f \\ 
                       && 4p$^{5}$5s$^{2}$   & 4p$^{5}$5s5p \\
                       && 4p$^{5}$5s5d   & 4p$^{5}$4d6p \\
                       && 4p$^{5}$4f6p & 4p$^{5}$5p6p \\[0.3cm]
    Y {\sc ii} Model  && 3d$^{10}$4s$^{2}$4p$^{6}$5s$^{2}$ & 3d$^{10}$4s$^{2}$4p$^{6}$4d5s \\
     - 13 configs && 3d$^{10}$4s$^{2}$4p$^{6}$4d$^{2}$ & 3d$^{10}$4s$^{2}$4p$^{6}$5s5p \\
   {\sc grasp}$^0$ &&3d$^{10}$4s$^{2}$4p$^{6}$4d5p  & 3d$^{10}$4s$^{2}$4p$^{6}$5p$^{2}$ \\
    && 3d$^{10}$4s$^{2}$4p$^{6}$5s5d & 3d$^{9}$4s$^{2}$4p$^{6}$5s5d$^{2}$ \\
    && 3d$^{10}$4s$^{2}$4p$^{6}$5d$^{2}$ & 3d$^{10}$4s$^{2}$4p$^{5}$5d$^{3}$ \\
    &&3d$^{10}$4s$^{2}$4p$^{4}$4d5d$^{3}$ & 3d$^{10}$4s$^{2}$4p$^{6}$4d5d \\
    && 3d$^{10}$4s$^{2}$4p$^{6}$5p5d\\ [0.2cm]  
    
    \hline\\
    \end{tabular}
    \caption{The configurations included in the wavefunction expansions for the {\sc autostructure} model for  Sr II and the {\sc grasp}$^0$ model for Y II. }
    \label{Table1}
    \end{table}

Table \ref{Table1} lists the valence orbitals and configurations included in the wavefunction representations of the analagous Hydrogen-like Sr {\sc ii} and Helium-like Y {\sc ii} targets respectively. The Sr {\sc ii} model is built up from twenty-two orbitals up to and including n=8s (1s, 2s, 2p, 3s, 3p, 3d, 4s, 4p, 4d, 4f, 5s, 5p, 5d, 5f, 5g, 6s, 6p, 6d, 7s, 7p, 7d and 8s). A total of 22 configurations were included in the wavefunction expansion giving rise to 298 individual fine-structure levels. Configurations including single promotions from the 4p orbital into the higher n=5 and 6 shells were included to provide additional configuration interaction (CI) to improve the energy levels. For Y {\sc ii} the target model comprised twelve orbitals up to and including n=5d (1s, 2s, 2p, 3s, 3p, 3d, 4s, 4p, 4d, 5s, 5p, 5d) and 13 configurations which gave rise to 2642 fine-structure levels. Single promotions from the 3d shell added some additional CI to improve the energy separations. The model for Sr {\sc ii} was generated using {\sc autostructure} whereas the Y {\sc ii} model was generated in {\sc grasp}$^0$ as these optimally represented the systems under investigation. It should be noted that these target models were kept to a computationally manageable size in anticipation of the electron-impact excitation calculations to follow, as the dense matrix manipulation associated with the increased number of coupled channels scales to a cubed power in the collisional work.

\begin{figure} 
\centering
    \includegraphics[width = 0.8\linewidth]{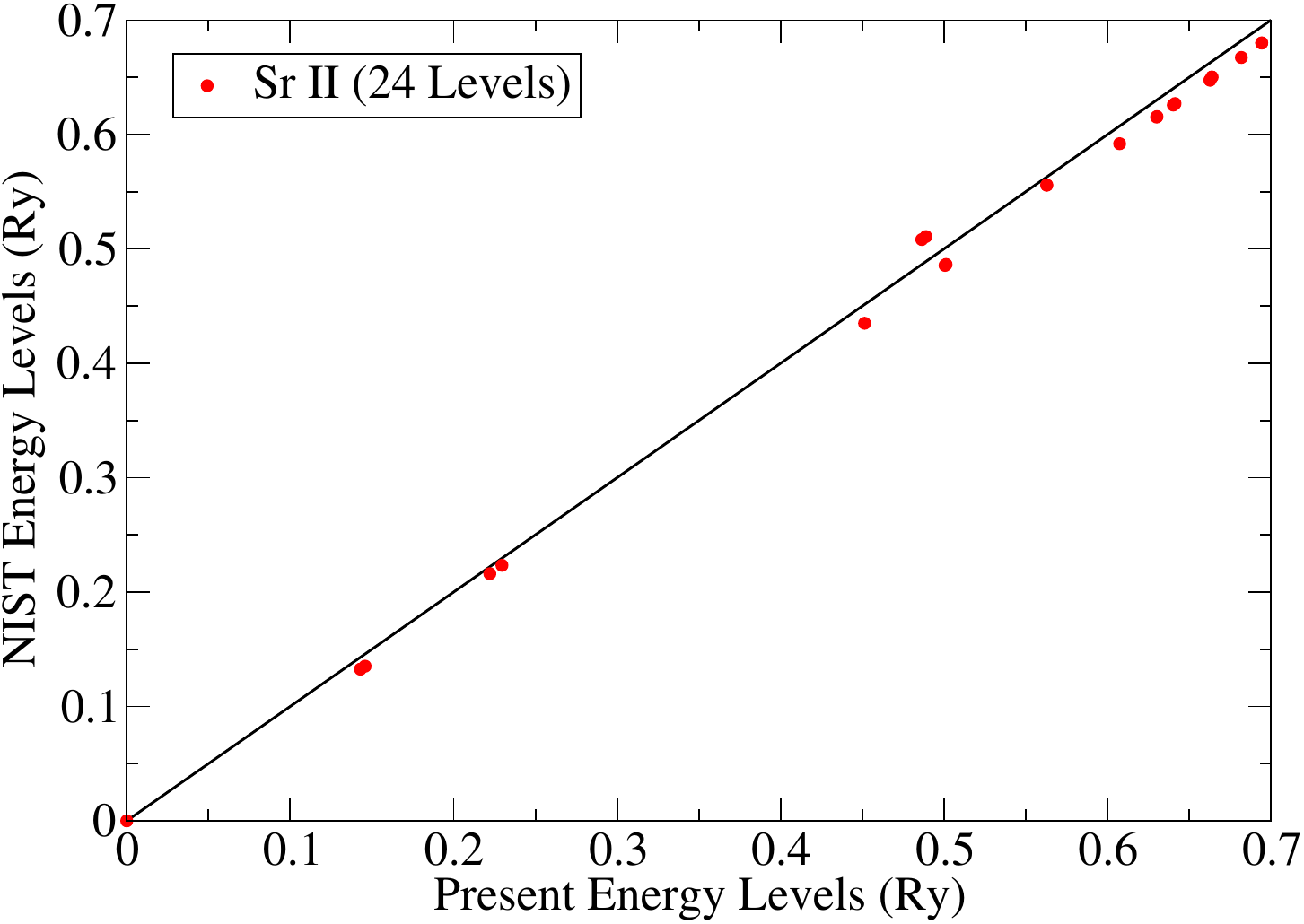}
    \caption{The lowest 24 energy levels in Ry computed using the
        present 22 configuration model for Sr {\sc ii} compared to the values
        available in the NIST database \citep{San12,Moore52}.}
    \label{Figure1}
\end{figure}

\begin{figure} 
\centering
    \includegraphics[width = 0.8\linewidth]{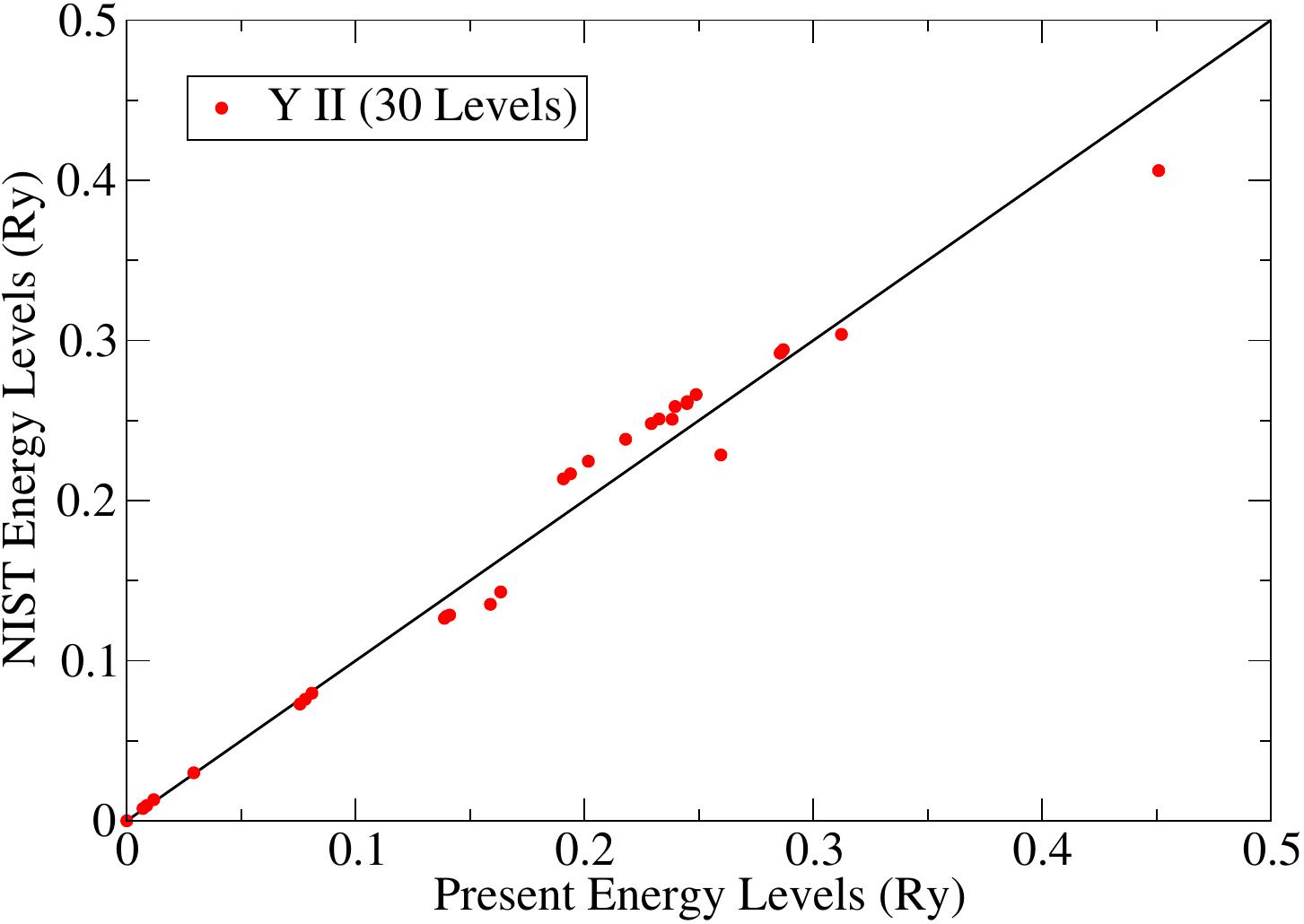}
    \caption{The lowest 30 energy levels in Ry computed using the
        present 13 configuration model for Y {\sc ii} compared to the values available in the NIST database \citep{Hannaford82}.}
    \label{Figure2}
\end{figure}

The target state energies for the first 24 levels of Sr {\sc ii} and the lowest 30 levels of Y {\sc ii} are presented in Tables \ref{Table2} and \ref{Table3} respectively. Comparisons are made with the data available in NIST \citep{nist} and the differences in Rydbergs ($1$ Ry $= 109,737.316$ {cm}$^{-1}$) listed. In Table 2 the majority of the Sr {\sc ii} energy separations are within approximately 0.01 Ry or less with the greatest difference of 0.02 Ry occurring for the 4p$^{6}$6p $^{2}$P$^{\rm{o}}_{3/2}$ and $^{2}$P$^{\rm{o}}_{5/2}$ levels indexed at levels 9 and 10. A similar picture emerges for the Y {\sc ii} target energies listed in Table 3 where the greatest disparities of approximately 0.03 Ry and 0.04 Ry are found for the 4d$^{2}$ $^{1}$S state (index 17) and the higher lying 5s5p $^{1}$P$^{\rm{o}}$ level (index 30) which was found to mix heavily with the lower lying 4d5p $^{1}$P$^{\rm{o}}$ state. The close conformity between predicted and observed energy levels for both systems is graphically displayed in Figure \ref{Figure1} for Sr {\sc ii} and Figure \ref{Figure2} for Y {\sc ii}. It should be noted that the thresholds included in both these target models including levels up to $\approx$ 0.5 - 0.6 Ry ($\approx$ 6.5 - 9 eV) cover the temperature range of interest in KNe modelling.

\begin{table}
    \centering
    \begin{tabular}{cclcccccccccc}
    \hline
    \\
    Index & Config           & Term                     & J   & NIST    & Present   & Diff           \\ 
          &                  &                          &     & Ry    & 22-config & Ry           \\ 
    \\
    \hline
    \\
    1     & $4$p$^{6}5$s & $^{2}$S$\phantom{\circ} $  & 0.5 & 0.00000 & 0.00000 & ---       \\       
    2     & $4$p$^{6}4$d & $^{2}$D$\phantom{\circ} $  & 1.5 & 0.13264 & 0.14294 & \phantom{-}0.0103 \\ 
    3     & $4$p$^{6}4$d & $^{2}$D$\phantom{\circ} $  & 2.5 & 0.13520 & 0.14581 & \phantom{-}0.0106  \\
    4     & $4$p$^{6}5$p & $^{2}$P$^{\circ}        $  & 0.5 & 0.21611 & 0.22215 & \phantom{-}0.0060  \\
    5     & $4$p$^{6}5$p & $^{2}$P$^{\circ}        $  & 1.5 & 0.22341 & 0.22957 & \phantom{-}0.0062  \\
    6     & $4$p$^{6}6$s & $^{2}$S$\phantom{\circ} $  & 0.5 & 0.43501 & 0.45144 & \phantom{-}0.0164\\       
    7     & $4$p$^{6}5$d & $^{2}$D$\phantom{\circ} $  & 1.5 & 0.48558 & 0.50065 & \phantom{-}0.0151\\
    8     & $4$p$^{6}5$d & $^{2}$D$\phantom{\circ} $  & 2.5 & 0.48637 & 0.50126 & \phantom{-}0.0149 \\        
    9     & $4$p$^{6}6$p & $^{2}$P$^{\circ}        $  & 0.5 & 0.50821 & 0.48645 &           -0.0218 \\
    10    & $4$p$^{6}6$p & $^{2}$P$^{\circ}        $  & 1.5 & 0.51084 & 0.48897 &           -0.0219 \\
    11    & $4$p$^{6}4$f & $^{2}$F$^{\circ}        $  & 3.5 & 0.55578 & 0.56287 & \phantom{-}0.0071 \\        
    12    & $4$p$^{6}4$f & $^{2}$F$^{\circ}        $  & 2.5 & 0.55579 & 0.56287 & \phantom{-}0.0071 \\      
    13    & $4$p$^{6}7$s & $^{2}$S$\phantom{\circ} $  & 0.5 & 0.59200 & 0.60746 & \phantom{-}0.0155 \\      
    14    & $4$p$^{6}6$d & $^{2}$D$\phantom{\circ} $  & 1.5 & 0.61531 & 0.63008 & \phantom{-}0.0148\\ 
    15    & $4$p$^{6}6$d & $^{2}$D$\phantom{\circ} $  & 2.5 & 0.61568 & 0.63031 & \phantom{-}0.0146\\ 
    16    & $4$p$^{6}7$p & $^{2}$P$^{\circ}        $  & 0.5 & 0.62585 & 0.64034 & \phantom{-}0.0145\\
    17    & $4$p$^{6}7$p & $^{2}$P$^{\circ}        $  & 1.5 & 0.62711 & 0.64134 & \phantom{-}0.0142 \\
    18    & $4$p$^{6}5$f & $^{2}$F$^{\circ}        $  & 2.5	& 0.64760 & 0.66277 & \phantom{-}0.0152 \\
    19    & $4$p$^{6}5$f & $^{2}$F$^{\circ}        $  & 3.5	& 0.64760 & 0.66277 & \phantom{-}0.0152 \\
    20    & $4$p$^{6}5$g & $^{2}$G	                  & 3.5	& 0.65026 & 0.66396 & \phantom{-}0.0137 \\
    21    & $4$p$^{6}5$g & $^{2}$G	                  & 4.5	& 0.65026 & 0.66396 & \phantom{-}0.0137 \\
    22    & $4$p$^{6}8$s & $^{2}$S	                  & 0.5	& 0.66739 & 0.68195 & \phantom{-}0.0146 \\
    23    & $4$p$^{6}7$d & $^{2}$D	                  & 1.5	& 0.68000 & 0.69434 & \phantom{-}0.0143 \\
    24    & $4$p$^{6}7$d & $^{2}$D	                  & 2.5	& 0.68020 & 0.69445 & \phantom{-}0.0142 \\      
    \\
    \hline
    \end{tabular}
    \caption{The 24 lowest energy levels in Rydbergs for the present 22 configuration {\sc autostructure} model compared to the values available in the NIST Database \citep{San12,Moore52}. The differences are presented in Rydbergs.}
    \label{Table2}
    \end{table}

\begin{table}
    \centering
    \begin{tabular}{cclcccccccccc}
    \hline
    \\
    Index & Config           & Term                     & J   & NIST    & Present   & Diff           \\ 
          &                  &                          &     & Ry    & 13-config & Ry           \\ 
    \\
    \hline
    \\
    1     & $5$s$^{2}$ & $^{1}$S$\phantom{\circ} $  & 0 & 0.0000 & 0.0000 & ---       \\       
    2     & $4$d$5$s   & $^{3}$D$\phantom{\circ} $  & 1 & 0.0077 & 0.0071 & \phantom{-}0.0006 \\ 
    3     & $4$d$5$s   & $^{3}$D$\phantom{\circ} $  & 2 & 0.0095 & 0.0087 & \phantom{-}0.0008  \\
    4     & $4$d$5$s   & $^{3}$D$\phantom{\circ} $  & 3 & 0.0132 & 0.0118 & \phantom{-}0.0014  \\
    5     & $4$d$5$s   & $^{1}$D$\phantom{\circ} $  & 2 & 0.0300 & 0.0293 & \phantom{-}0.0007  \\
    6     & $4$d$^{2}$ & $^{3}$F$\phantom{\circ} $  & 2 & 0.0729 & 0.0757 & -0.0028\\       
    7     & $4$d$^{2}$ & $^{3}$F$\phantom{\circ} $  & 3 & 0.0759 & 0.0780 & -0.0021 \\          
    8     & $4$d$^{2}$ & $^{3}$F$\phantom{\circ} $  & 4 & 0.0797 & 0.0809 & -0.0013 \\        
    9     & $4$d$^{2}$ & $^{3}$P$\phantom{\circ} $  & 0 & 0.1265 & 0.1388 & -0.0123 \\         
    10    & $4$d$^{2}$ & $^{3}$P$\phantom{\circ} $  & 1 & 0.1277 & 0.1397 & -0.0120 \\          
    11    & $4$d$^{2}$ & $^{3}$P$\phantom{\circ} $  & 2 & 0.1285 & 0.1411 & -0.0126 \\        
    12    & $4$d$^{2}$ & $^{1}$D$\phantom{\circ} $  & 2 & 0.1352 & 0.1589 & -0.0238 \\      
    13    & $4$d$^{2}$ & $^{1}$G$\phantom{\circ} $  & 4 & 0.1429 & 0.1634 & -0.0205 \\      
    14    & $5$s$5$p   & $^{3}$P$^{\circ} $         & 0 & 0.2136 & 0.1908 & \phantom{-}0.0229\\ 
    15    & $5$s$5$p   & $^{3}$P$^{\circ} $         & 1 & 0.2167 & 0.1939 & \phantom{-}0.0227\\ 
    16    & $5$s$5$p   & $^{3}$P$^{\circ} $         & 2 & 0.2246 & 0.2017 & \phantom{-}0.0229\\
    17    & $4$d$^{2}$ & $^{1}$S$\phantom{\circ} $  & 0 & 0.2285 & 0.2596 & -0.0312 \\      
    18    & $4$d$5$p   & $^{1}$D$^{\circ} $         & 2 & 0.2383 & 0.2180 & \phantom{-}0.0203\\ 
    19    & $4$d$5$p   & $^{3}$F$^{\circ} $         & 2 & 0.2481 & 0.2292 & \phantom{-}0.0189\\ 
    20    & $4$d$5$p   & $^{1}$P$^{\circ} $         & 1 & 0.2508 & 0.2383 & \phantom{-}0.0125\\ 
    21    & $4$d$5$p   & $^{3}$F$^{\circ} $         & 3 & 0.2509 & 0.2326 & \phantom{-}0.0183\\ 
    22    & $4$d$5$p   & $^{3}$F$^{\circ} $         & 4 & 0.2587 & 0.2396 & \phantom{-}0.0191\\ 
    23    & $4$d$5$p   & $^{3}$D$^{\circ} $         & 1 & 0.2606 & 0.2448 & \phantom{-}0.0158\\ 
    24    & $4$d$5$p   & $^{3}$D$^{\circ} $         & 2 & 0.2618 & 0.2449 & \phantom{-}0.0169\\ 
    25    & $4$d$5$p   & $^{3}$D$^{\circ} $         & 3 & 0.2662 & 0.2488 & \phantom{-}0.0174\\ 
    26    & $4$d$5$p   & $^{3}$P$^{\circ} $         & 0 & 0.2921 & 0.2855 & \phantom{-}0.0065\\   
    27    & $4$d$5$p   & $^{3}$P$^{\circ} $         & 1 & 0.2927 & 0.2860 & \phantom{-}0.0067\\  
    28    & $4$d$5$p   & $^{3}$P$^{\circ} $         & 2 & 0.2942 & 0.2869 & \phantom{-}0.0072\\ 
    29    & $4$d$5$p   & $^{1}$F$^{\circ} $         & 3 & 0.3038 & 0.3123 & \phantom{-}0.0085\\ 
    30    & $5$s$5$p   & $^{1}$P$^{\circ}$          & 1 & 0.4061 & 0.4509 & -0.0448\\ 
    \\
    \hline
    \end{tabular}
    \caption{The 30 lowest energy levels in Rydbergs for the present 13 configuration {\sc grasp0} model compared to the values available in the NIST Database \citep{Nilsson91}. The differences are presented in Rydbergs.}
    \label{Table3}
    \end{table}


A further test to gauge the accuracy of the target state wavefunction representations presented above for Sr {\sc ii} and Y {\sc ii} is to compare known values of the Einstein A-values for transitions among the target levels. The evaluation of these transition rates are dependent on accurate energy separations as any disparities in the transition energy $\Delta E$ are scaled by ($\Delta E$)$^3$ for the dipole allowed E1 and M1 transitions and ($\Delta E$)$^5$ for quadrupole lines. 
To alleviate this, the current A-values are recalculated using the spectroscopically accurate NIST values listed in Tables \ref{Table2} and \ref{Table3} according to,

\begin{eqnarray}
A_{j\to i}(\mathrm{Shifted)} = 
   \Big( \frac{ \Delta E_{\text{NIST}} }{ \Delta E_{\text{Calculated}} }\Big )^3A_{j\to i}(\mathrm{Unshifted}),
\end{eqnarray}
with the quadrupoles similarly multiplied by the ratio to the power 5. This ensures that the A-values are computed using energies which have been shifted to their spectroscopically accurate energy values.

\begin{figure} 
\centering
    \includegraphics[width = 0.9\linewidth]{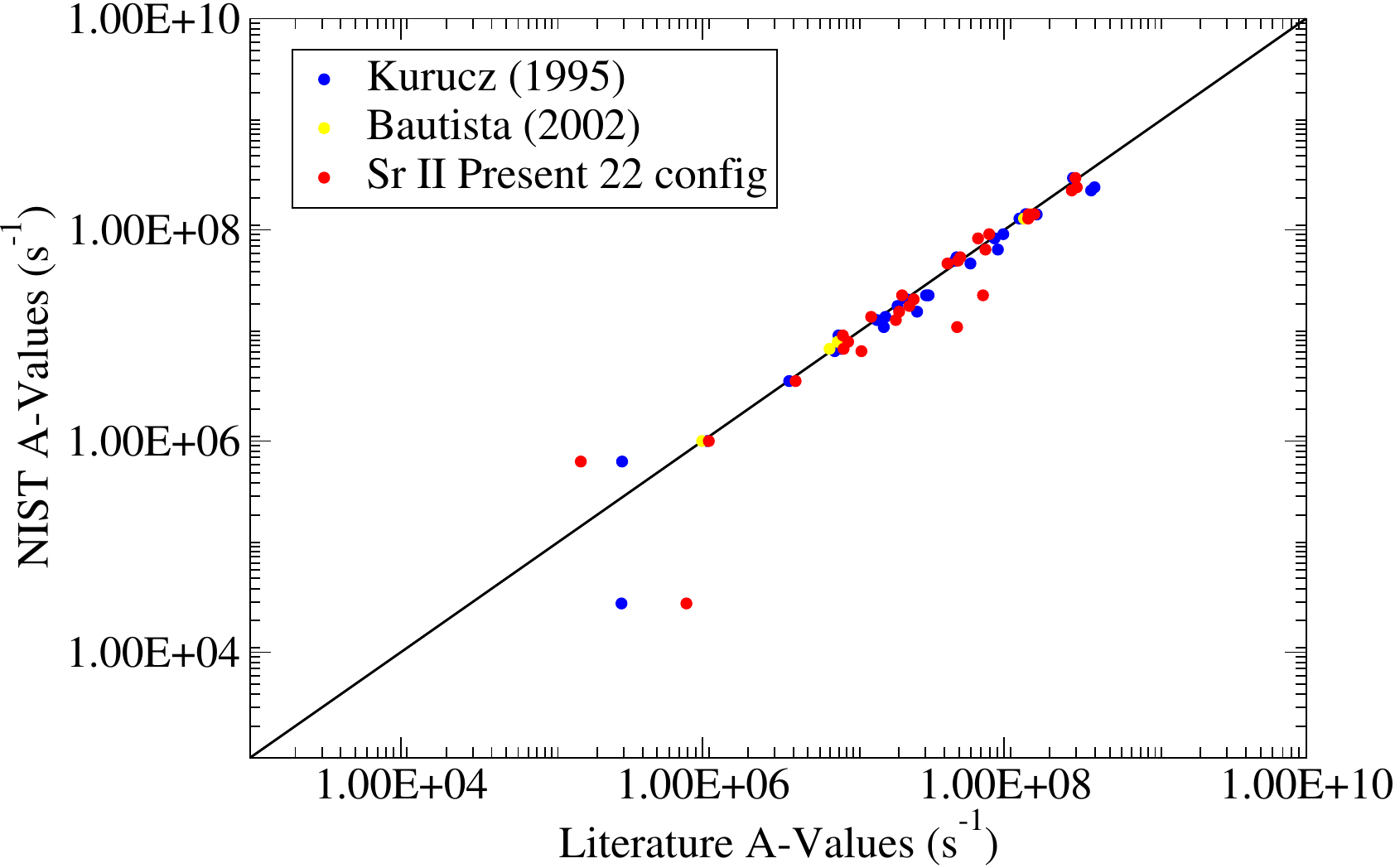}
    \caption{Einstein A-values for all E1 and M1 dipole transitions among the
        lowest 24 levels of Sr {\sc ii} computed by the present 22-configuration
        {\sc autostructure} model (red), those available in the \citet{Kurucz1995} database (blue) and those computed by \citet{Bautista2002} (yellow), compared to the values available in the NIST database \citep{WieMar80}.}
    \label{Figure3}
\end{figure}

\begin{figure} 
\centering
    \includegraphics[width = 0.9\linewidth]{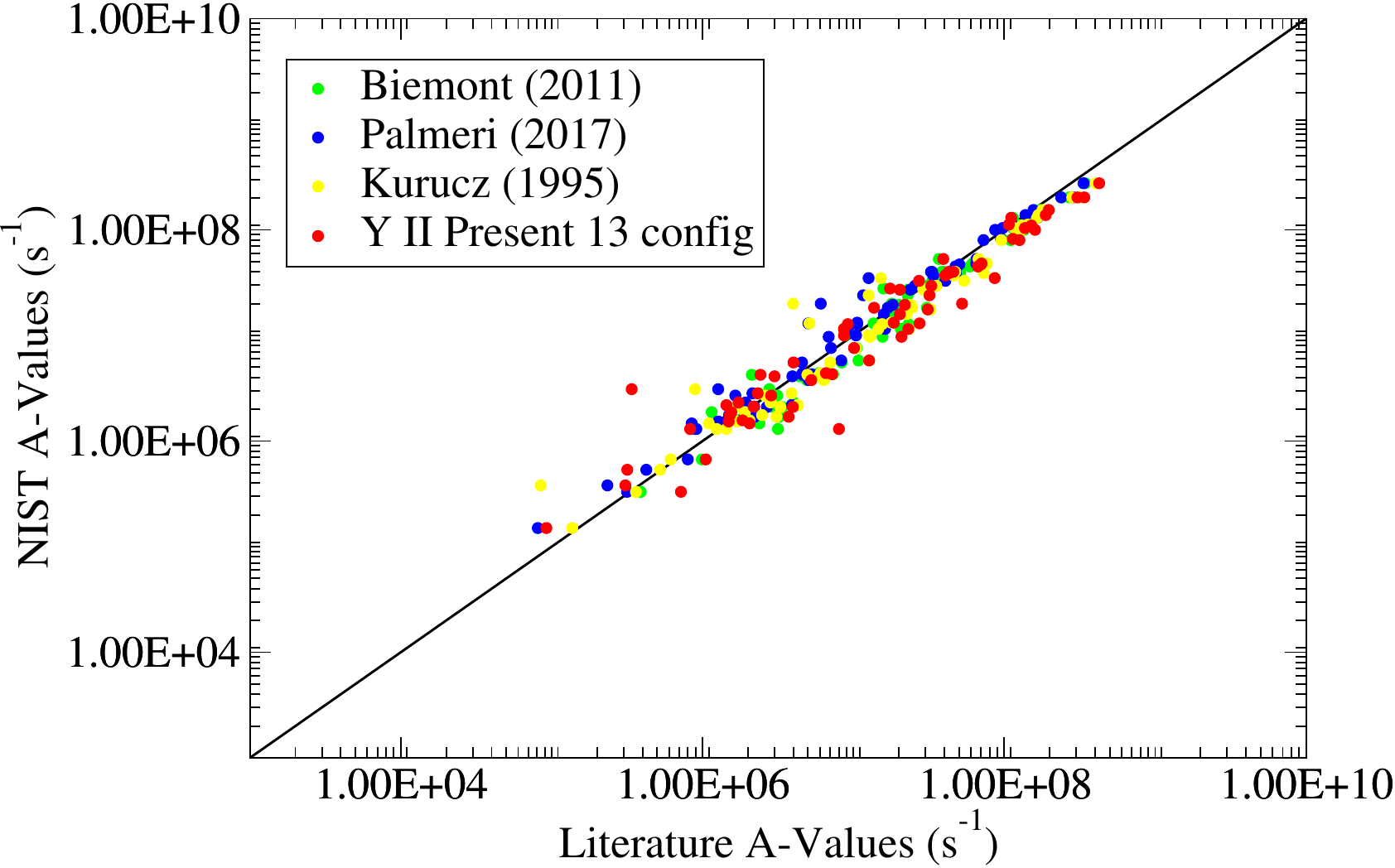}
    \caption{Einstein A-values for all E1 and M1 dipole transitions among the
            lowest 30 levels of Y {\sc ii} computed by, the present 13-configuration {\sc grasp0} model (red), \citet{Biemont11} (green), \citet{Palmeri17} (blue), \citet{Kurucz1995} (yellow), compared to the values available in the NIST database \citep{Hannaford82}.}
    \label{Figure4}
\end{figure}
\begin{table*}
\centering
    \begin{tabular}{crrccccccc}
    \hline \\
         Sr {\sc ii} lines &&&&&& \vspace{0.5mm} \\
$\lambda$ &Index & $E_{{i}}$\phantom{00} & Lower & $E_{{j}}$ & Upper\phantom{-} & 
\multicolumn{4}{c}{$A_{{j\to i}}$ (s$^{-1}$)}\\  

(nm)&(${i}$-${j}$)\phantom{0} & (cm$^{-1}$) &  Level ${i}$ & (cm$^{-1}$) & Level ${j}$ & NIST & Present & Kur95 & Baut02\\
\\
\hline
\\
\phantom{-}407.89  & 1-5 & 0.000 &  4p$^{6}$5s $^{2}$S$_{1/2}$ &  24516.65 &  4p$^{6}$5p $^{2}$P$^{\circ}_{3/2}$ & 1.41E+08   & 1.58E+08 &  1.41E+08 & 1.50E+08\\
\phantom{-}416.30  & 4-6 & 23715.19 &  4p$^{6}$5p $^{2}$P$^{\circ}_{1/2}$ & 47736.53 &  4p$^{6}$6s $^{2}$S$_{1/2}$ & 6.50E+07 & 7.49E+07 &  6.52E+07 & -\\
\phantom{-}421.67  & 1-4 & 0.000 &  4p$^{6}$5s $^{2}$S$_{1/2}$ &  23715.19 &  4p$^{6}$5p $^{2}$P$^{\circ}_{1/2}$ & 1.28E+08   & 1.44E+08 &  1.26E+08 & 1.35E+08\\
\phantom{-}430.67  & 5-6 & 24516.65 &  4p$^{6}$5p $^{2}$P$^{\circ}_{3/2}$ & 47736.53 &  4p$^{6}$6s $^{2}$S$_{1/2}$ & 1.40E+08 & 1.45E+08 &  1.40E+08 & -\\
\phantom{-}474.37  & 6-17 & 47736.53 &  4p$^{6}$6s $^{2}$S$_{1/2}$ & 68817.12 &  4p$^{6}$7p $^{2}$P$^{\circ}_{3/2}$ & -       & 2.12E+06 &  8.93E+05 & -\\
\phantom{-}477.49  & 6-16 & 47736.53 &  4p$^{6}$6s $^{2}$S$_{1/2}$ & 68679.34 &  4p$^{6}$7p $^{2}$P$^{\circ}_{1/2}$ & -       & 1.65E+06 &  8.77E+05 & -\\
\phantom{-}964.53  & 11-20 & 60990.04 &  4p$^{6}$4f $^{2}$F$^{\circ}_{7/2}$ & 71357.80 &  4p$^{6}$5g $^{2}$G$_{7/2}$ & -      & 2.70E+06 &  2.64E+06 & -\\
\phantom{-}964.53  & 11-21 & 60991.34 &  4p$^{6}$4f $^{2}$F$^{\circ}_{7/2}$ & 71357.80 &  4p$^{6}$5g $^{2}$G$_{9/2}$ & -      & 7.58E+07 &  7.40E+07 & -\\
\phantom{-}964.65  & 12-20 & 60991.34 &  4p$^{6}$4f $^{2}$F$^{\circ}_{5/2}$ & 71357.80 &  4p$^{6}$5g $^{2}$G$_{7/2}$ & -      & 7.30E+07 &  7.13E+07 & -\\
1003.94 & 2-5 & 14555.90 & 4p$^{6}$4d $^{2}$D$_{3/2}$ & 24516.65  & 4p$^{6}$5p $^{2}$P$^{\circ}_{3/2}$  & 1.00E+06            & 1.10E+06 &  9.97E+05 & 9.93E+05 \\
1033.01 & 3-5 & 14836.24 & 4p$^{6}$4d $^{2}$D$_{5/2}$ & 24516.65  & 4p$^{6}$5p $^{2}$P$^{\circ}_{3/2}$ &  8.70E+06            & 9.22E+06 &  8.79E+06 & 7.88E+06 \\
1087.62 & 9-13 & 55769.70 & 4p$^{6}$6p $^{2}$P$^{\circ}_{1/2}$ & 64964.10 & 4p$^{6}$7s $^{2}$S$_{1/2}$ & -                    & 1.87E+07 &  1.99E+07 & - \\
1091.79 & 2-4 & 14555.90 & 4p$^{6}$4d $^{2}$D$_{3/2}$ & 23715.19 &  4p$^{6}$5p $^{2}$P$^{\circ}_{1/2}$  &  7.46E+06  &          8.61E+06 &  7.46E+06 & 6.94E+06 \\
1122.81 & 10-13 & 56057.90 & 4p$^{6}$6p $^{2}$P$^{\circ}_{3/2}$ & 64964.10 & 4p$^{6}$7s $^{2}$S$_{1/2}$ & -                  &  3.56E+07 &  3.61E+07 & - \\
\\
\hline
\end{tabular}
\caption{Representative sample of Sr II spectral lines. The transition probabilities $A_{j \rightarrow i}$ calculated here are compared with those available on the NIST database \citep{nist,reader1980_nist_lines,brage1998theoretical_nist_lines} and the calculations of \citet{Kurucz1995,Bautista2002}.}
\label{tab:srii_lines}
\end{table*}
    
\begin{table*}
\centering
    \begin{tabular}{crrcccccccc}
        \hline
        \vspace{2mm}\\
        Y {\sc ii} lines &&&&&& \vspace{0.5mm}
        \\
        $\lambda$ &Index & $E_{{i}}$\phantom{00} & Lower & $E_{{j}}$ & Upper\phantom{-} & 
        \multicolumn{5}{c}{$A_{{j\to i}}$ (s$^{-1}$)}\\  
        
        (nm)&(${i}$-${j}$)\phantom{0} & (cm$^{-1}$) &Level ${i}$ &   (cm$^{-1}$) & Level ${j}$ & NIST & Present & Kur95 & Bie11 & Pal17 \\
        \\
        \hline
        \\
        540.43 & 12 - 29 & 14832.86 &  4d$^2$ $^1$D$_{2}$ & 33336.72 & 4d5p $^1$F$_{3}$  &  - &  1.91E+07 &  1.42E+07 & 1.54E+07 &  1.50E+07 \\               
        547.49 & 10 - 28 & 14018.27 &  4d$^2$ $^3$P$_{1}$ & 32283.42 & 4d5p $^3$P$_{2}$  &  4.30E+06 &  7.22E+06 &  4.25E+06 & 7.34E+06 &  5.28E+06 \\                      
        548.23 &  9 - 27 & 13883.38 &  4d$^2$ $^3$P$_{0}$ & 32124.05 & 4d5p $^3$P$_{1}$  &  7.60E+06 &  1.01E+07 &  7.57E+06 & 1.05E+07 &  7.10E+06 \\                      
        549.89 & 11 - 28 & 14098.07 &  4d$^2$ $^3$P$_{2}$ & 32283.42 & 4d5p $^3$P$_{2}$  &  1.15E+07 &  2.31E+07 &  1.16E+07 & 2.12E+07 &  1.62E+07 \\                      
        551.14 &  6 - 18 &  8003.13 &  4d$^2$ $^3$F$_{2}$ & 26147.25 & 4d5p $^1$D$_{2}$  &  4.24E+06 &  2.42E+06 &  4.29E+06 & 2.12E+06 &  4.60E+06 \\                      
        554.62 & 10 - 26 & 14018.27 &  4d$^2$ $^3$P$_{1}$ & 32048.79 & 4d5p $^3$P$_{0}$  &  1.76E+07 &  3.11E+07 &  1.76E+07 & 3.20E+07 &  2.22E+07 \\                      
        554.75 & 11 - 27 & 14098.07 &  4d$^2$ $^3$P$_{2}$ & 32124.05 & 4d5p $^3$P$_{1}$  &  5.80E+06 &  1.27E+07 &  5.74E+06 & 1.08E+07 &  8.30E+06 \\                      
        566.45 & 13 - 29 & 15682.90 &  4d$^2$ $^1$G$_{4}$ & 33336.72 & 4d5p $^1$F$_{3}$  &  - &  7.27E+07 &  7.19E+07 & 6.57E+07 &  6.31E+07 \\               
        661.56 & 11 - 25 & 14098.07 &  4d$^2$ $^3$P$_{2}$ & 29213.96 & 4d5p $^3$D$_{3}$  &  1.70E+06 &  3.72E+06 &  1.69E+06 & 3.22E+06 &  2.21E+06 \\                      
        679.72 &  9 - 23 & 13883.38 &  4d$^2$ $^3$P$_{0}$ & 28595.28 & 4d5p $^3$D$_{1}$  &  - &  1.15E+06 &  1.74E+06 & 1.41E+06 &  1.20E+06 \\               
        679.73 & 10 - 24 & 14018.27 &  4d$^2$ $^3$P$_{1}$ & 28730.00 & 4d5p $^3$D$_{2}$  &  - &  2.73E+06 &  2.51E+06 & 2.70E+06 &  1.73E+06 \\               
        683.44 & 11 - 24 & 14098.07 &  4d$^2$ $^3$P$_{2}$ & 28730.00 & 4d5p $^3$D$_{2}$  &  3.30E+05 &  7.19E+05 &  3.28E+05 & 3.89E+05 &  3.16E+05 \\                      
        686.01 & 10 - 23 & 14018.27 &  4d$^2$ $^3$P$_{1}$ & 28595.28 & 4d5p $^3$D$_{1}$  &  - &  7.97E+05 &  9.02E+05 & 9.70E+05 &  6.67E+05 \\               
        689.79 & 11 - 23 & 14098.07 &  4d$^2$ $^3$P$_{2}$ & 28595.28 & 4d5p $^3$D$_{1}$  &  - &  3.20E+05 &  1.52E+05 & 9.84E+05 &  2.01E+05 \\               
        726.62 & 12 - 23 & 14832.86 &  4d$^2$ $^1$D$_{2}$ & 28595.28 & 4d5p $^3$D$_{1}$  &  1.30E+06 &  8.01E+06 &  1.33E+06 & 3.16E+06 &  8.93E+05 \\                      
        733.50 &  9 - 20 & 13883.38 &  4d$^2$ $^3$P$_{0}$ & 27516.69 & 4d5p $^1$P$_{1}$  &  - &  6.66E+05 &  5.52E+04 & 4.33E+05 &  3.90E+04 \\               
        745.23 & 11 - 20 & 14098.07 &  4d$^2$ $^3$P$_{2}$ & 27516.69 & 4d5p $^1$P$_{1}$  &  - &  3.59E+04 &  3.63E+05 & 1.45E+06 &  6.80E+05 \\               
        788.41 & 12 - 20 & 14832.86 &  4d$^2$ $^1$D$_{2}$ & 27516.69 & 4d5p $^1$P$_{1}$  &  1.00E+07 &  8.66E+06 &  9.63E+06 & 8.74E+06 &  1.04E+07 \\                      
       3033.82 &  1 -  5 &     0.00 &  5s$^2$ $^1$S$_{0}$ &  3296.18 & 5s4d $^1$D$_{2}$  &  - &  8.44E-04 &  9.49E-04 & - &  - \\                    
       4071.69 &  2 -  5 &   840.20 &    5s4d $^3$D$_{1}$ &  3296.18 & 5s4d $^1$D$_{2}$  &  - &  4.20E-03 &  5.72E-03 & - &  - \\            
       4442.27 &  3 -  5 &  1045.08 &    5s4d $^3$D$_{2}$ &  3296.18 & 5s4d $^1$D$_{2}$  &  - &  5.94E-04 &  8.05E-04 & - &  - \\     
       \\
       \hline
    \end{tabular}
    \caption{Representative sample of Y II spectral lines. The transition probabilities $A_{j \rightarrow i}$ calculated here are compared with those available on the NIST database \citep{nist,Hannaford82} and the calculations of 
    \citet{Kurucz1995,Biemont11,Palmeri17} .}
    \label{tab:yii_lines}
\end{table*}

\begin{table}
{%
\begin{tabular}{@{} c c c c @{}}
\hline 
\\
Index & Config  &  $\tau$ Present (ns) &  $\tau$ Expt (ns) 
\\
\hline
\\
\multicolumn{4}{l}{Sr {\sc ii}}
\vspace{1mm}\\
2  & $4$p$^{6}4$d  $^{2}$D$        _{3/2}    $  & 3.82$^8$ & (4.35$\pm$0.04)$^8$$^{a}$                         \\
   &                                            &          & (4.35$\pm$0.04)$^8$$^{b}$                          \\
3  & $4$p$^{6}4$d  $^{2}$D$        _{5/2}    $  & 3.44$^8$ & (3.72$\pm$0.25)$^8$$^{c}$                        \\ 
   &                                            &          & (4.08$\pm$0.22)$^8$$^{b}$                          \\
4  & $4$p$^{6}5$p  $^{2}$P$^{\circ}_{1/2}    $  & 6.55     & 7.39 $\pm$ 0.07$^{d}$                  \\
5  & $4$p$^{6}5$p  $^{2}$P$^{\circ}_{3/2}    $  & 5.94     & 6.63 $\pm$ 0.07$^{d}$                   \\
11 & $4$p$^{6}4$f  $^{2}$F$^{\circ}_{7/2}    $  & 3.05     & 2.97 ± 0.05$^{d}$        \\
12 & $4$p$^{6}4$f  $^{2}$F$^{\circ}_{5/2}    $  & 3.07     & 3.09 ± 0.06$^{d}$        \\
\\
\multicolumn{4}{l}{Y {\sc ii}}
\vspace{1mm}\\                           
26 & $4$d$5$p    $^{3}$P$^{\circ}_0$ & 2.19 & 2.8$\pm$0.2$^{e}$                                                                                   \\
27 & $4$d$5$p    $^{3}$P$^{\circ}_1$ & 2.20 & 2.8$\pm$0.2$^{e}$                                                                                   \\
28 & $4$d$5$p    $^{3}$P$^{\circ}_2$ & 2.23 & 2.6$\pm$0.2$^{e}$                                                                                   \\
29 & $4$d$5$p    $^{1}$F$^{\circ}_3$ & 4.32 & 4.7$\pm$0.3$^{e}$                                                                                   \\
30 & $5$s$5$p    $^{1}$P$^{\circ}_1$ & 0.96 & 1.2$\pm$0.2$^{e}$                                                                                   \\
31 & $4$d$5$d    $^{1}$F$_3$ & 2.41     & 2.43$\pm$0.10$^{f}$                                                                                     \\
32 & $4$d$5$d    $^{3}$D$_1$ & 2.60     & 2.60$\pm$0.15$^{f}$                                                                                     \\
33 & $5$p$^{2}$  $^{1}$F$_0$ & 1.25     & 1.77$\pm$0.09$^{f}$                                                                                     \\
34 & $4$d$5$d    $^{3}$D$_2$ & 2.64     & 2.53$\pm$0.10$^{f}$                                                                                     \\
35 & $5$p$^{2}$  $^{3}$P$_1$ & 1.24     & 1.92$\pm$0.10$^{f}$                                                                                     \\
36 & $4$d$5$d    $^{3}$G$_3$ & 2.15     & 2.53$\pm$0.15$^{f}$                                                                                     \\
37 & $4$d$5$d    $^{3}$D$_3$ & 2.62     & 2.64$\pm$0.15$^{f}$                                                                                     \\
38 & $4$d$5$d    $^{3}$G$_4$ & 2.15     & 2.45$\pm$0.15$^{f}$                                                                                     \\
39 & $5$p$^{2}$  $^{3}$P$_2$ & 1.54     & 2.29$\pm$0.10$^{f}$                                                                                     \\
40 & $4$d$5$d    $^{1}$P$_1$ & 2.75     & 2.64$\pm$0.10$^{f}$                                                                                     \\
41 & $5$p$^{2}$  $^{3}$G$_5$ & 2.20     & 2.59$\pm$0.10$^{f}$                                                                                     \\
42 & $4$d$5$d    $^{1}$D$_2$ & 2.59     & 4.36$\pm$0.20$^{f}$                                                                                     \\
50 & $4$d$5$d    $^{3}$P$_1$ & 1.93     & 1.30$\pm$0.07$^{f}$                                                                                     \\
51 & $4$d$5$d    $^{3}$P$_2$ & 1.91     & 1.23$\pm$0.05$^{f}$                                                                                     \\ \bottomrule
\end{tabular}%
}
\caption{The radiative lifetimes $\tau$ (ns) for some selected states of Sr {\sc ii} and Y {\sc ii}. The superscripts denote the attributed sources of experimental values $a$ - \protect\cite{Mannervik1999}, $b$ -  \protect\cite{Bimont2000}, $c$ - \protect\cite{Madej1990},$d$ -  \protect\cite{Pinnington1995}, $e$ - \protect\cite{Biemont11},$f$ -  \protect\cite{Palmeri17}. ()$^{8}$ denotes the value to be in standard notation of $\times 10^{8}$}
\label{tab:lifetimes}
\end{table}

In Figures \ref{Figure3} and \ref{Figure4} we present a comparison of all known literature A-values for the E1 and M1 allowed lines among the lowest 24 levels of Sr {\sc ii} and the lowest 30 levels of Y {\sc ii}. Evidently there is a paucity of available data for the Sr {\sc ii} ion with only five lines published by \cite{Bautista2002} and the only other available data comes from the \cite{Kurucz1995} database from their { gfemq3801.all} and {gfemq3901.all} repositories. For the strongest lines, however, there is reasonable agreement between the theoretical predictions with the main outliers coming from the weaker lines. It should also be noted that the accuracy of the A-values given in the NIST database range from AA - D (2-50\%) for many of these lines. The situation for Y {\sc ii} is much better with A-value computations from \cite{Biemont11}, \cite{Palmeri17} and \cite{Kurucz1995} available for comparison. Figure \ref{Figure4} depicts a better linear representation for the E1/M1 comparison with NIST showing reasonable conformity among all the theoretical predictions. The accuracy of these A-values is very important in any modelling, whether LTE or NLTE. 

To investigate this further we present in Table \ref{tab:srii_lines} a selection of strong dipole important lines which have been identified as useful in astrophysical modelling. For Sr {\sc ii} we concentrate on the observed lines in Eta Carinae discussed by \cite{Bautista2002}, which originate from resonant emission produced by transitions among the first five levels, 4p$^{6}$5s, 4p$^{6}$4d and 4p$^{6}$5p. In addition, to identify absorption features in the X-Shooter spectra of the KNe AT2017gfo \cite{Watson19}, sought weak lines, blueshifted by 0.1 - 0.3c, in the two wavelength windows 390 - 500 nm and 900 - 1160 nm in the rest frame. Sr {\sc ii} features were found at restframe wavelengths between 1000 - 1100 nm, the most prominent of which was centred close to 1050 nm. All (allowed) transitions that lie within these wavelength windows are tabulated in Table \ref{tab:srii_lines} and excellent agreement is found between all datasets. For the case of Y {\sc ii} the lines chosen are the strongest E1 lines that fall in the mid to near ultraviolet (200 - 400 nm) as well as those listed in the AT2017gfo P Cygni line analysis work of \cite{SnepWat23} These strong dipole transitions are among the 4d$^{2}$ and 4d5p levels in the visible with a mean wavelength in the range 760 - 770 nm (LTE-weighted) with the most prominent feature located at 788.19 nm. There is also the possibility of Y {\sc ii} line identifications occurring at longer wavelengths, possibly in the near to intermediate infrared, but this has yet to be confirmed by any modelling. The lines of interest are for transitons between 4d$^{2}$ levels and the higher lying 4d5s and 4d5p states. The strongest of these lines lie in the wavelength range 800 nm to 4 $\mu$m and are included in the A-value tabulations for Y {\sc ii} in Table \ref{tab:yii_lines}. The agreement between all five datasets is excellent for the majority of the transitions listed, a few outliers exhibit a more satisfactory agreement but it is not always consistent as to which theoretical model produces the anomaly.

As a final test to confirm the accuracy of these target models, we consider the radiative lifetimes of the target states of both Sr {\sc ii} and Y {\sc ii}.  In Table \ref{tab:lifetimes} we present the radiative lifetimes for a selection of Sr {\sc ii} and Y {\sc ii} states and compare with a comprehensive set of recent experimental measurements from the literature. This list is not exhaustive and only those states with reasonably long lived lifetimes which have been measured experimentally are considered. The agreement between theory and experiment for all states presented gives additional confidence in the theoretical models adopted in this work.

\section{Electron-Impact Excitation Collision Calculations}
The electron-impact excitation calculations presented in this paper were computed within the framework of the close-coupling $R$-matrix method, a detailed description of which can be found in \cite{Burke} and will not be repeated here. In summary $R$-matrix theory divides configuration space into two distinct regions, the inner and outer regions. These regions are separated by an $R$-matrix boundary at $r=a$, which is chosen to completely enclose the most diffuse orbital and hence the charge distribution of the $N$-electron target. This boundary thus acts as an interface between the two regions. The $R$-matrix is defined as,

\begin{equation}
    R_{ij}=\frac{1}{2a}\sum_k^{N+1}\frac{\omega_{ik}(a)\omega_{jk}(a)}{E_k^{N+1}-E},
\end{equation}
where $E_k^{N+1}$ are the eigenenergies of the ($N+1$) Hamiltonian, $E$ is the energy of the incident electron, and $\omega_{ik}$ are the surface amplitudes. In the internal region, electron exchange and short range correlation effects between the incident electron and the target are strong and cannot be neglected. In the external region the free electron moves only in the long range potential of the target and hence electron exchange and correlation effects can be ignored. The suite of $R$-matrix packages allows for the computation of the electron-impact collision strengths, $\Omega_{i\to j}$, for excitation from some initial level $i$ to some final level $j$. These collision strengths are related to the cross section $\sigma_{i\rightarrow j}$ by the relation,
\begin{equation}
    \Omega_{i\rightarrow j}=\frac{g_ik_i^2}{\pi a_0^2}\sigma_{i\rightarrow j},
\end{equation}
where $g_i$ is the statistical weight of the initial state, $k_i^2$ is the energy of the incident electron in Rydbergs, and $a_0$ is the Bohr radius. Effective collision strengths ($\Upsilon_{ij}$) can then be evaluated by Maxwellian averaging over a Boltzmann distribution of electron temperatures so that,
\begin{equation}
    \Upsilon_{ij}(T_e)=\int^\infty_0\Omega_{i\rightarrow j}e^{-\epsilon_j/kT_e}d \left(\frac{\epsilon_j}{kT_e}\right),
\end{equation}
where the scattered electron has residual energy $\epsilon_j$. The terms $k$ and $T_e$ are Boltzmann's constant and the electron temperature (in Kelvin), respectively. It is these Maxwellian averaged effective collision strengths that are commonly used by astrophysical and plasma modellers in their diagnostic applications.

There are several variants of the $R$-matrix computer codes currently available for use in collision investigations, two of which are used in the current calculations. The Sr {\sc ii} work was performed by incorporating the 22-configuration model from {\sc autostructure} into the semi-relativistic Breit-Pauli suite of codes \citep{Badnell86,Badnell97} where the expansion of the target wavefunction is constructed in intermediate coupling and the spin–orbit operator is included in the ($N$ + 1)-electron Hamiltonian. For Y {\sc ii} the fully relativistic {\sc pdarc} (Parallel Dirac Atomic $R$-matrix Code) was adopted which permits a fully relativistic $jj$ coupled scattering calculation to be performed by solving the Dirac equation with the Dirac Hamiltonian \cite{Norrington04} and \cite{Norrington87}. All of these computer packages are freely available at \cite{Ballance}.

\begin{figure*}
    \centering
    \includegraphics[width=0.78 \linewidth]{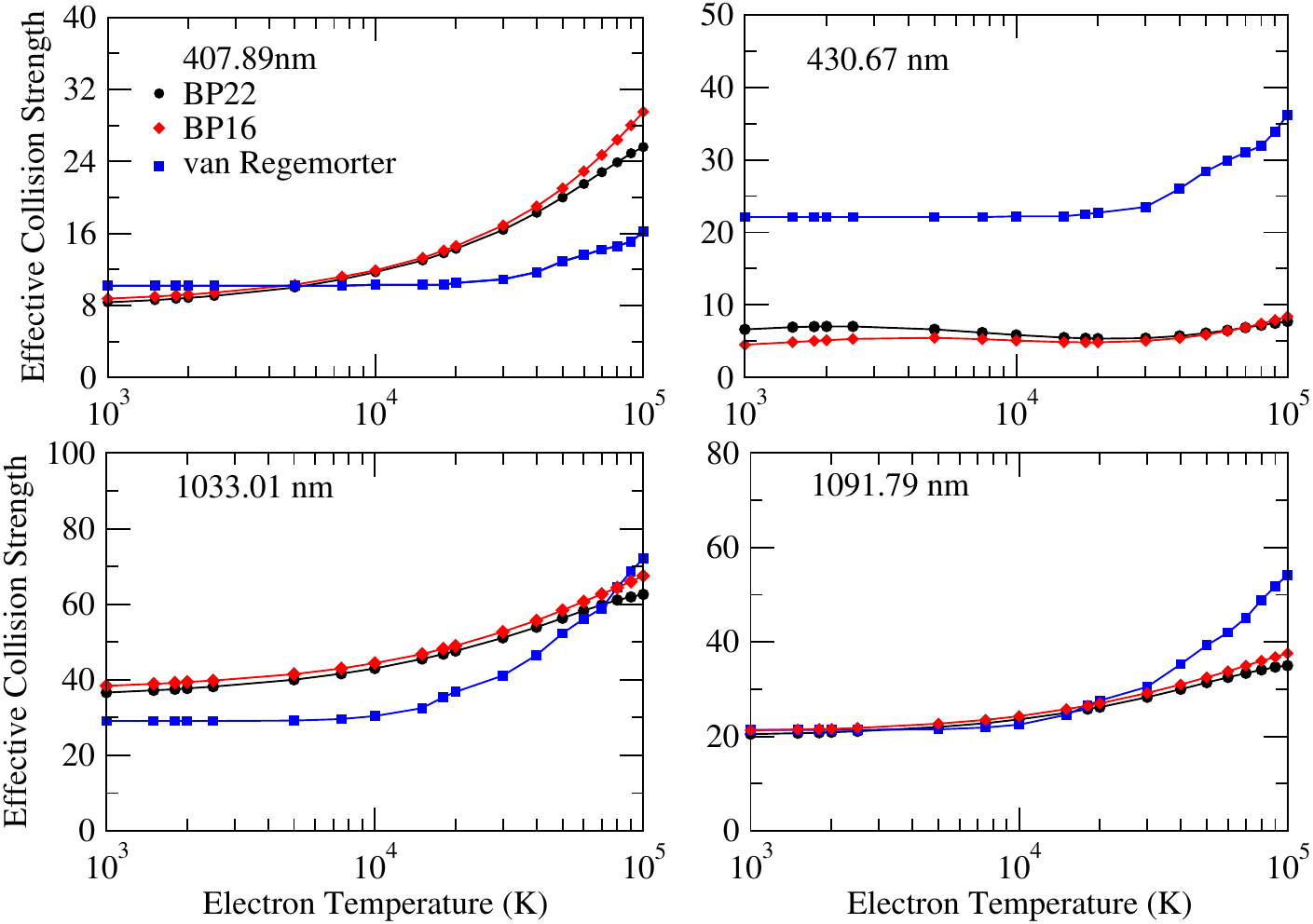}
    \caption{Effective collision strengths for four Sr II E1 transitions selected from Table \ref{tab:srii_lines}. The transitions shown are 5s $^2$S$_{1/2}$ $\to$ 5p $^2$P$^{\mathrm{o}}_{3/2}$ ($1\to5, \lambda = 407.89$ nm),   5p $^2$P$^{\mathrm{o}}_{3/2}$ $\to$ 6s $^2$S$_{1/2}$ ($5\to6, \lambda = 430.67$ nm), 4d $^2$D$_{5/2}$ $\to$ 5p $^2$P$^{\mathrm{o}}_{3/2}$ ($3\to5, \lambda = 1033.01$ nm) and 4d $^2$D$_{3/2}$ $\to$ 5p $^2$P$^{\mathrm{o}}_{1/2}$ ($2\to4, \lambda = 1091.79$ nm). Two RMBP calculations are shown with 22 and 16 NRCSFs respectively. Additionally, the approximation techniques of \citet{VanReg1962} is shown. }
    \label{fig:sr2_ecs}
\end{figure*}

\begin{figure*}
    \centering
    \includegraphics[width=0.78\linewidth]{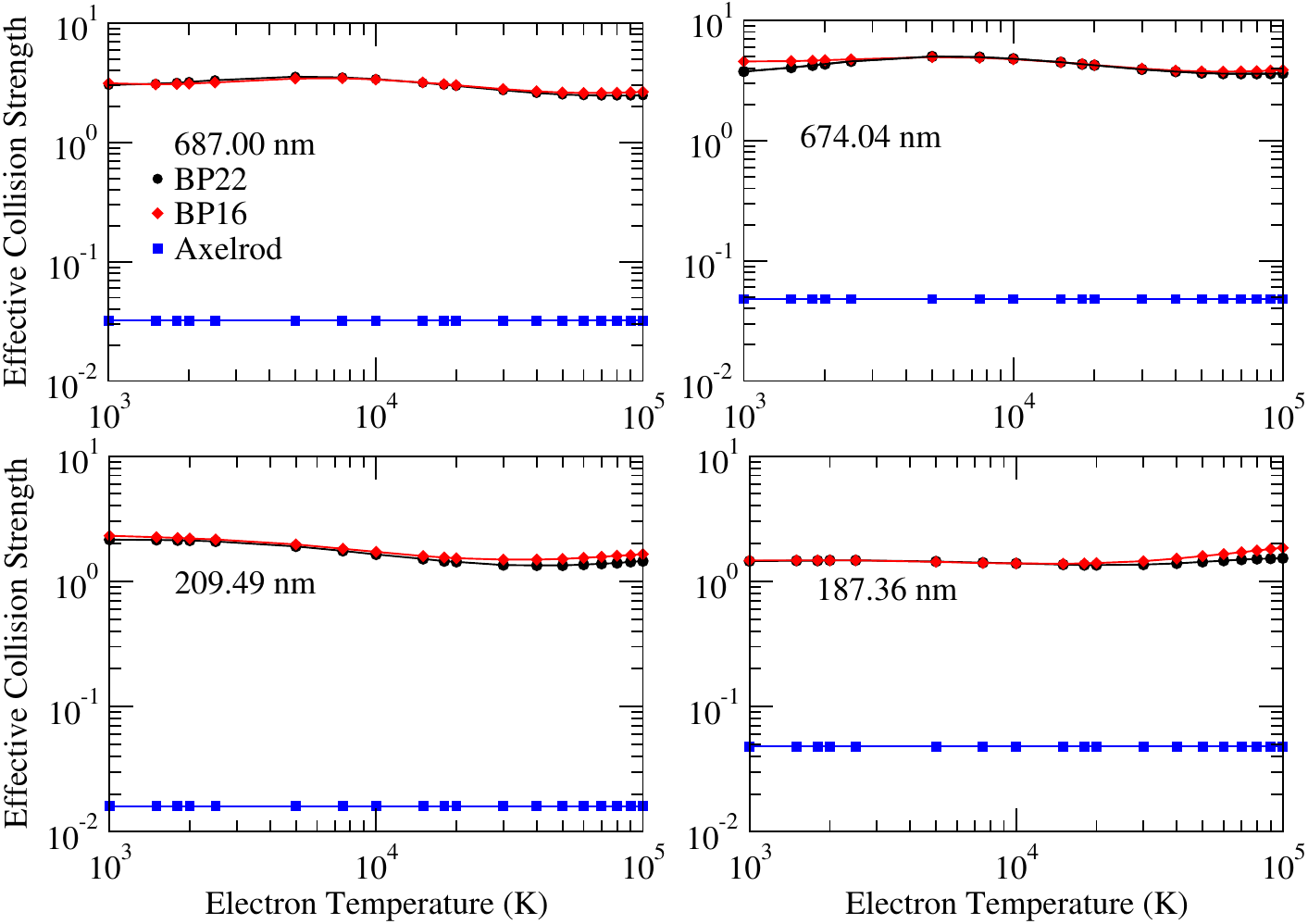}
    \caption{Effective collison strengths for four Sr II forbidden transitions selected from Table \ref{tab:srii_lines}. The transitions shown are 5s $^2$S$_{1/2}$ $\to$ 4d $^2$D$_{3/2}$ ($1\to2, \lambda = 687.00$ nm),   5s $^2$S$_{1/2}$ $\to$ 4d $^2$D$_{5/2}$ ($1\to3, \lambda = 674.04$ nm), 5s $^2$S$_{1/2}$ $\to$ 6s $^2$S$_{1/2}$ ($1\to6, \lambda = 209.49$ nm) and 5s $^2$S$_{1/2}$ $\to$  5d $^2$D$_{5/2}$ ($1\to8, \lambda = 187.36$ nm). Two RMBP calculations are shown with 22 and 16 NRCSFs respectively. Additionally, the approximation techniques of \t{Axelrod1980} is shown.}
    \label{fig:sr2_ecs_forbidden}
\end{figure*}

\begin{figure*}
    \centering
    \includegraphics[width=0.80\linewidth]{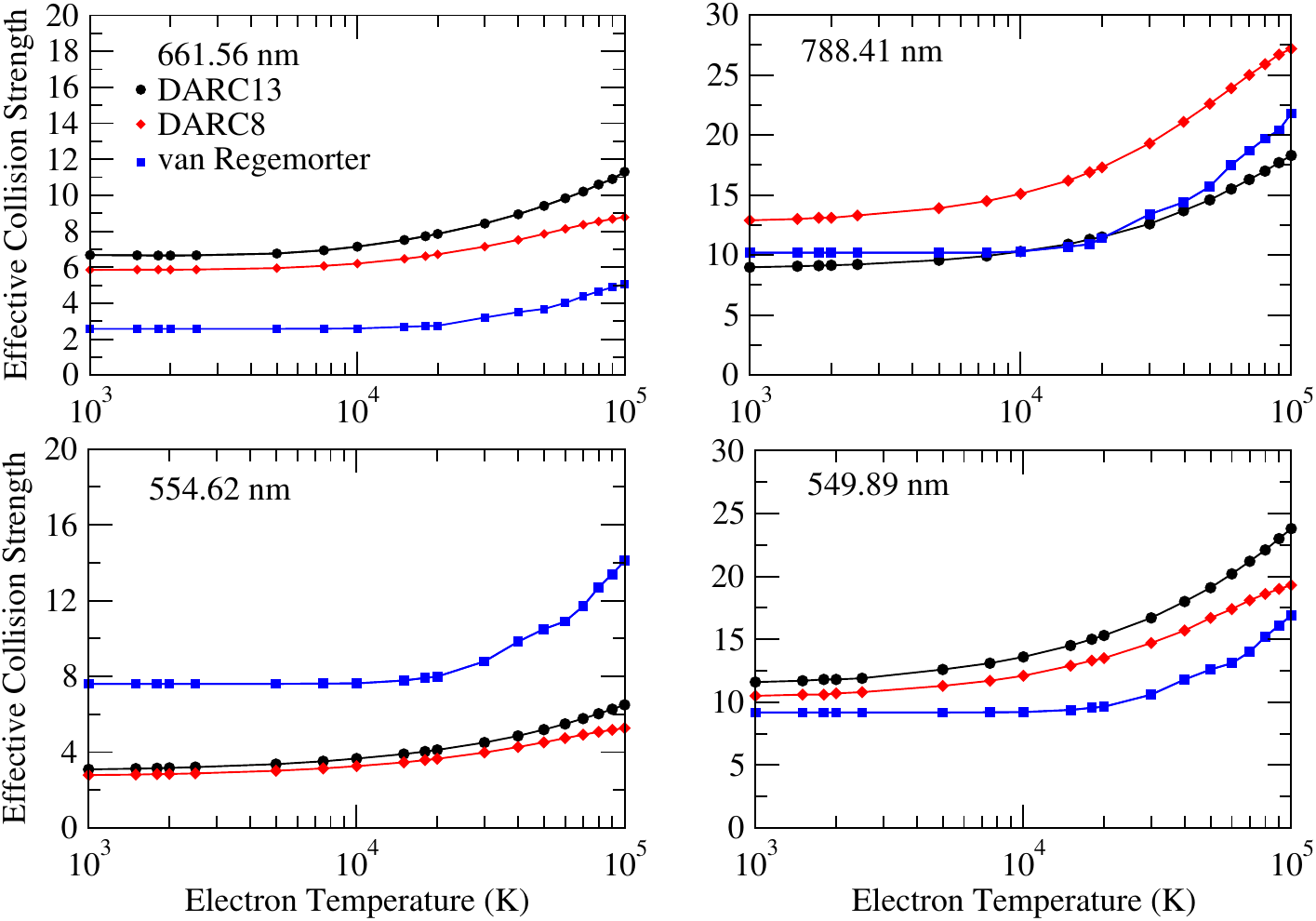}
    \caption{Effective collison strengths for four Y II E1 transitions selected from Table \ref{tab:yii_lines}. The transitions shown are 4d$^2$ $^3$P$_{2}$ $\to$ 4d5p $^3$D$^{\mathrm{o}}_{3}$ ($11\to25, \lambda = 661.56$ nm),   4d$^2$ $^1$D$_{2}$ $\to$ 4d5p $^1$P$^{\mathrm{o}}_{1}$ ($12\to20, \lambda = 788.41$ nm), 4d$^2$ $^3$P$_{1}$ $\to$ 4d5p $^3$S$^{\mathrm{o}}_{0}$ ($10\to26, \lambda = 554.62$ nm) and 4d$^2$ $^3$P$_{2}$ $\to$ 4d5p $^3$P$^{\mathrm{o}}_{2}$ ($11\to28, \lambda = 549.89$ nm). }
    \label{fig:y2_ecs}
\end{figure*}

\begin{figure*}
    \centering
    \includegraphics[width=0.81\linewidth]{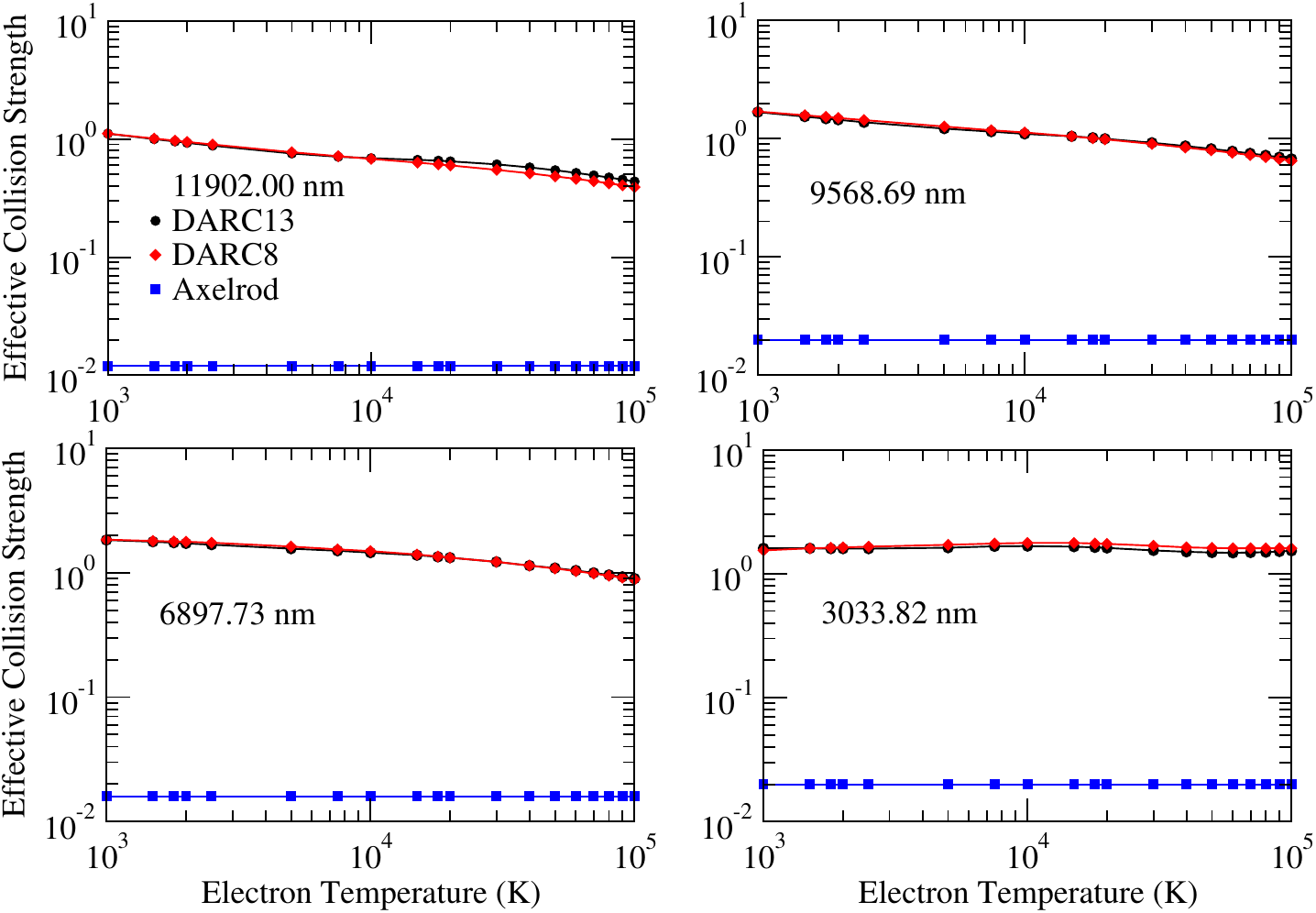}
    \caption{Effective collison strengths for four Y II forbidden transitions selected from Table \ref{tab:yii_lines}. The transitions shown are 5s$^2$ $^1$S$_{0}$ $\to$ 5s4d $^3$D$_{1}$ ($1\to2, \lambda = 11902.00$ nm),   5s$^2$ $^1$S$_{0}$ $\to$ 5s4d $^3$D$_{2}$ ($1\to3, \lambda = 9568.69$ nm), 5s$^2$ $^1$S$_{0}$ $\to$ 5s4d $^3$D$_{3}$ ($1\to4, \lambda = 6897.73$ nm) and 5s$^2$ $^1$S$_{0}$ $\to$ 5s4d $^1$D$_{2}$ ($1\to5, \lambda = 3033.82$ nm).}
    \label{fig:y2_ecs_forbidden}
\end{figure*}

\begin{table*}
\centering
    \begin{tabular}{cccccccccccccccccccccc}
    \hline
    & \multicolumn{10}{c}{Temperature (K)}\\
             &  1.00E+03 & 2.00E+03 & 5.00E+03 & 1.00E+04 & 2.00E+04 & 4.00E+04 & 5.00E+04 & 7.00E+04 & 8.00E+04 & 1.00E+05 \\
    \hline
    \\
    Index & \multicolumn{10}{c}{Effective Collision Strengths}\\
    \\
      1-2 & 3.03E+00 & 3.22E+00 & 3.56E+00 & 3.40E+00 & 2.98E+00 & 2.60E+00 & 2.52E+00 & 2.47E+00 & 2.47E+00 & 2.48E+00 \\
      1-3 & 3.75E+00 & 4.33E+00 & 5.00E+00 & 4.79E+00 & 4.23E+00 & 3.74E+00 & 3.64E+00 & 3.59E+00 & 3.59E+00 & 3.62E+00 \\
      1-4 & 4.57E+00 & 4.73E+00 & 5.26E+00 & 6.08E+00 & 7.44E+00 & 9.57E+00 & 1.05E+01 & 1.19E+01 & 1.25E+01 & 1.34E+01 \\
      1-5 & 8.35E+00 & 8.84E+00 & 1.00E+01 & 1.17E+01 & 1.43E+01 & 1.83E+01 & 2.00E+01 & 2.28E+01 & 2.39E+01 & 2.56E+01 \\
      2-3 & 1.36E+01 & 1.42E+01 & 1.49E+01 & 1.38E+01 & 1.17E+01 & 9.36E+00 & 8.63E+00 & 7.60E+00 & 7.22E+00 & 6.64E+00 \\
      2-4 & 2.05E+01 & 2.09E+01 & 2.20E+01 & 2.36E+01 & 2.62E+01 & 3.00E+01 & 3.14E+01 & 3.34E+01 & 3.41E+01 & 3.50E+01 \\
      2-5 & 6.61E+00 & 6.75E+00 & 7.08E+00 & 7.48E+00 & 7.85E+00 & 8.16E+00 & 8.28E+00 & 8.46E+00 & 8.51E+00 & 8.54E+00 \\
      3-4 & 4.57E+00 & 3.80E+00 & 3.11E+00 & 2.85E+00 & 2.59E+00 & 2.19E+00 & 2.05E+00 & 1.85E+00 & 1.78E+00 & 1.66E+00 \\
      3-5 & 3.66E+01 & 3.77E+01 & 4.00E+01 & 4.30E+01 & 4.76E+01 & 5.39E+01 & 5.63E+01 & 5.98E+01 & 6.11E+01 & 6.26E+01 \\
      4-5 & 5.30E+00 & 5.34E+00 & 5.51E+00 & 5.84E+00 & 6.09E+00 & 6.11E+00 & 6.11E+00 & 6.15E+00 & 6.18E+00 & 6.21E+00 \\
    \\
    \hline
    \end{tabular}
    \label{tab:srii_ecs} 
    \caption{Effective collision strengths as a function of electron temperature for transitions between the first five fine structure levels of Sr II. A full dataset will be made available at \citet{openadas_site}.}
\end{table*}

\begin{table*}
\centering
   \begin{tabular}{cccccccccccccccccccccc}
    \hline
    & \multicolumn{10}{c}{Temperature (K)}\\
             &  1.00E+03 & 2.00E+03 & 5.00E+03 & 1.00E+04 & 2.00E+04 & 4.00E+04 & 5.00E+04 & 7.00E+04 & 8.00E+04 & 1.00E+05 \\
    \hline
    \\
    Index & \multicolumn{10}{c}{Effective Collision Strengths }\\
    \\
    1-2  & 1.18E+00  & 9.66E-01 & 7.64E-01 & 6.69E-01 & 5.96E-01 & 5.15E-01 & 4.87E-01 & 4.43E-01 & 4.26E-01 & 3.96E-01 \\ 
    1-3  & 1.69E+00  & 1.45E+00 & 1.21E+00 & 1.09E+00 & 9.72E-01 & 8.39E-01 & 7.94E-01 & 7.27E-01 & 6.99E-01 & 6.53E-01 \\ 
    1-4  & 1.88E+00  & 1.75E+00 & 1.57E+00 & 1.45E+00 & 1.31E+00 & 1.14E+00 & 1.08E+00 & 9.89E-01 & 9.51E-01 & 8.88E-01 \\ 
    1-5  & 1.64E+00  & 1.70E+00 & 1.72E+00 & 1.76E+00 & 1.72E+00 & 1.61E+00 & 1.59E+00 & 1.58E+00 & 1.59E+00 & 1.63E+00 \\
    2-3  & 7.38E+00  & 6.88E+00 & 6.11E+00 & 5.41E+00 & 4.58E+00 & 3.74E+00 & 3.50E+00 & 3.18E+00 & 3.07E+00 & 2.89E+00 \\
    2-4  & 4.22E+00  & 4.18E+00 & 4.03E+00 & 3.77E+00 & 3.32E+00 & 2.74E+00 & 2.56E+00 & 2.28E+00 & 2.17E+00 & 1.99E+00 \\
    2-5  & 3.14E+00  & 2.98E+00 & 2.84E+00 & 2.79E+00 & 2.62E+00 & 2.24E+00 & 2.10E+00 & 1.88E+00 & 1.80E+00 & 1.66E+00 \\
    3-4  & 1.04E+01  & 1.02E+01 & 9.60E+00 & 8.83E+00 & 7.71E+00 & 6.41E+00 & 6.00E+00 & 5.42E+00 & 5.20E+00 & 4.86E+00 \\
    3-5  & 5.14E+00  & 4.91E+00 & 4.62E+00 & 4.38E+00 & 3.93E+00 & 3.33E+00 & 3.14E+00 & 2.85E+00 & 2.73E+00 & 2.55E+00 \\
    4-5  & 6.95E+00  & 6.55E+00 & 6.03E+00 & 5.61E+00 & 4.98E+00 & 4.25E+00 & 4.03E+00 & 3.69E+00 & 3.55E+00 & 3.33E+00 \\
    \\
    \hline
    \end{tabular}
    \caption{Effective collision strengths as a function of electron temperature for transitions between the first five fine structure levels of Y II. A full dataset will be made available at \citet{openadas_site}.}
    \label{tab:yii_ecs}
\end{table*}

\subsection{Sr \sc{ii}}

The 22 configuration model for Sr {\sc ii} gave rise to 298 individual fine structure levels of which the lowest 24 were shifted to their spectroscopic positions compiled by NIST to ensure that the thresholds were at their correct positions aiding the identification of lines for spectral analysis. A total of 30 continuum orbitals were included for each channel angular momentum and the $R$-matrix boundary radius was set at 61.9 a.u.
The $R$-matrix calculations were carried out for all partial waves with total angular momentum 2$J$ $\leq$ 62 and the ($N$+1)-Hamiltonian matrices had a maximum size of 51258 $\times$ 51258. For the forbidden transitions these parameters are sufficient to ensure convergence of the corresponding collision strengths. For the dipole allowed transitions, however, it is necessary to include higher partial waves 2$J$ $>$ 62. To estimate the contribution to the collision strength from these higher partial waves a Bethe sum-rule (\cite{Burgess1974}) is adopted for the dipole transtions and for quadrupole and higher it is assumed that the partial collision strengths form a geometric series as a function of partial wave. Finally the high energy Bethe and Born limits were computed to check the validity of the collision strengths at higher energies.
A fine mesh of 12800 points with an energy spacing of 7.81E-05 Ry was used for all partial waves $2J$ $\leq$ 14 to fully resolve the resonance structure in the collision cross sections for energies $E$ $\leq$ 1.0 Ry. For incident electron energies above this, 1.0 < $E$ $\leq$ 2.0 Ry, a coarser mesh of 1280 points and an energy spacing of 7.81E-04 Ry was adopted. Finally, the infinite energy points for dipole lines were calculated from the Einstein A-coefficients as described by \cite{Burgess1992analysis} and for the non-dipoles the Born approximation discussed by \citep{EISSNER1991} was applied.

We present in Figure \ref{fig:sr2_ecs} the Maxwellian-averaged effective collision strengths as a function of electron temperature (K), for four strong E1 dipole transitions previously highlighted in Table \ref{tab:srii_lines}. The lines chosen are the 407.89 nm (5s $^{2}$S$_{1/2}$ - 5p $^{2}$P$^{\circ}_{3/2}$ 1-5), the 430.67 nm (5p $^{2}$P$^{\circ}_{3/2}$ - 6s $^{2}$S$_{3/2}$ 5-6), the 1033.01 nm (4d $^{2}$D$_{5/2}$ - 5p $^{2}$P$^{\circ}_{3/2}$ 3-5) and the 1091.79 nm (4d $^{2}$D$_{3/2}$ - 5p $^{2}$P$^{\circ}_{1/2}$ 2-4). These transitions 
 are representative strong lines in the visible or the near-infrared, reflecting the most interesting regions of the KNe spectrum. The NIR lines in the bottom frames of the figure have been previously highlighted by \cite{Watson19} as a Sr {\sc ii} identification in the KNe spectrum. The atomic data associated with these lines will therefore prove vital in any subsequent NLTE modelling of KNe and the conclusive identification of spectral features. There is currently no other data available in the literature with which to compare these effective collision strengths so as an accuracy crosscheck we have computed the corresponding data from a smaller 16 configuration Breit-Pauli model. For all four dipole transitions presented excellent agreement between the two calculations is evident at all temperatures considered. This provides confidence in the data presented and indicates convergence has been reached for the effective collision strengths presented. 

Approximate excitation rates can also be computed using the \cite{VanReg1962} formula which expresses the effective collision strength as,
\begin{equation}
       \Upsilon_{ij} = (2.39\text{E+06}) P(y) \lambda^3 A_{j \rightarrow i} g_i, \label{eq:vr} 
\end{equation}

where $\lambda$ is the wavelength in cm,  $y = {E_{ij}} / kT$ with $E_{ij}$ the transition energy and $P(y)$ is a tabulated function for singly-ionised species shown in \cite{VanReg1962} which is interpolated for the requested temperature range. This approximation has been used for allowed (electric dipole, E1) transitions in calculations made by large scale modelling codes such as {\sc sumo} \citep{pognan2022} and {\sc artis} \citep{shingles2020}. For additional comparison, we also show in Figure \ref{fig:sr2_ecs} 
the effective collision strengths computed using this approximation for all four transitions considered and a mixed bag of results is evident. In general, the van Regemorter formula gives results of the correct order of magnitude and temperature dependence but appears to either under or overestimate the effective collision strength on a transition by transition basis. 

For the forbidden transitions such codes employ a further approximation developed by \cite{Axelrod1980}, so that
\begin{equation}
    \Upsilon_{ij} = 0.004 g_i g_j \label{eq:axel},
\end{equation}
where $g_i$ and $g_j$ are the statistical weights of the lower and upper levels respectively and the formula is notably independent of temperature. In Figure \ref{fig:sr2_ecs_forbidden} we present the comparison for four forbidden lines among the lowest lying levels of Sr {\sc ii}, namely the 687.00 nm (5s $^2$S$_{1/2}$ - 4d $^2$D$_{3/2}$, 1-2) , 674.04 nm (5s $^2$S$_{1/2}$ - 4d $^2$D$_{5/2}$, 1-3), 209.49 nm (5s $^2$S$_{1/2}$ - 4d $^2$S$_{1/2}$, 1-6) and 187.36 nm (5s $^2$S$_{1/2}$ - 5d $^2$D$_{5/2}$, 1-8) spectral lines.

Clearly the Axelrod approximation systematically underestimates the effective collision strengths by more than two orders of magnitude for all temperatures considered. 
These approximations are commonly used procedures in large scale modelling codes for the mass production of such excitation data and given the comparisons discussed above caution on their use is advised, particularly for forbidden lines.

\subsection{Y II}

For the case of Y {\sc ii} the 13 configuration model consisted of a total of 2642 individual fine structure levels but only the lowest 30 were retained in the close-coupling expansion of the scattering wavefunction. The retention of these low lying levels was sufficient for the KNe modelling of interest. Similar to the Sr {\sc ii} case, all 30 levels were shifted to their spectroscopic positions observed by NIST to enable accurate line identification. The $R$-matrix boundary was set at 18.08 a.u, 25 continuum orbitals were included for each channel angular momentum, the $R$-matrix calculations were carried out for all partial waves with total angular momentum 2J $\leq$ 79 and contributions to the collision strength from higher partial waves 2J $>$ 79 were included as described for Sr {\sc ii}. A fine mesh of 25000 points with an energy spacing of 1.0E-04 Ry was used to fully resolve the resonance structure in the collision cross sections across the energy range of interest 0-2.5 Ry.

We present in Figure \ref{fig:y2_ecs} the Maxwellian-averaged effective collision strengths as a function of electron temperature (K), for four strong E1 dipole transitions previously highlighted in Table \ref{tab:yii_lines}. The lines chosen are the 661.56 nm (4d$^{2}$ $^{3}$P$_{2}$ - 4d5p $^{3}$D$^{\circ}_{3}$ 11-25), the 788.41 nm (4d$^{2}$ $^{1}$D$_{2}$ - 4d5p $^{1}$P$^{\circ}_{1}$ 12-20), the 554.62 nm (4d$^{2}$ $^{3}$P$_{1}$ - 4d5p $^{3}$S$^{\circ}_{0}$ 10-26) and the 549.89 nm (4d$^{2}$ $^{3}$P$_{2}$ - 4d5p $^{3}$P$^{\circ}_{2}$ 11-28). 
The top panel lines (661.56 and 788.41 nm) were chosen as they were 
listed in the AT2017gfo P Cygni line analysis work of \cite{SnepWat23}
and represent strong dipole transitions prominent in the visible wavelength region. The lower panel transitions (554.62 and 549.89 nm) were chosen as they were found to be two strong E1 lines responsible for the most transitions, as a fraction of all transitions, between 400-800 nm when we perform LTE \textsc{tardis} modelling of the kilonova as 100\% Y in Section 4 of this paper. As was the case for Sr {\sc ii} there is currently no other data available in the literature with which to compare these effective collision strengths so we again compare with the corresponding data from a smaller 8 configuration fully relativistic DARC model. Good agreement between the two calculations is evident at all temperatures considered with the exception of the 788.41 nm line where differences of on average 20\% are found across the temperatures of interest. On investigation it was found that the upper level for this transition , conventionally labeled as 4d5p $^{1}$P$^{\circ}_{1}$ (level 20), shows a significant contribution in the corresponding eigenvector from the 5s5p $^{1}$P$^{\circ}_{1}$ configuration. In the smaller 8 configuration DARC model the $jj$-mixing purity was 57\% whereas for the 13 configuration model the purity of the upper state increased to 80\% . This different percentage mixing of the levels along with the differences in the corresponding A-values for this transition leads to the 20\% disparty between the two models. The effective collision strengths produced by the van Regemorter approximation are plotted for comparison, underestimating for the 661.56 nm line, overestimating for the 554.62 nm line and producing satisfactory values for the other two lines. Finally in Figure \ref{fig:y2_ecs_forbidden} we present the comparison for four forbidden lines among the lowest lying levels of Y {\sc ii}, namely the 11902.00 nm (5s$^2$ $^1$S$_0$ - 4d5s $^3$D$_1$, 1-2) , 9568.69 nm (5s$^2$ $^1$S$_0$ - 4d5s $^3$D$_2$, 1-3), 6897.73 nm (5s$^2$ $^1$S$_0$ - 4d5s $^3$D$_3$, 1-4) and 3033.82 nm (5s$^2$ $^1$S$_0$ - 4d5s $^1$D$_2$, 1-5) spectral lines.
Again we find that the Axelrod approximation systematically underestimates the effective collision strengths by more than two orders of magnitude for all temperatures considered.

In Tables \ref{tab:srii_ecs} and \ref{tab:yii_ecs} we present the effective collision strengths for all transitions among the lowest 5 levels of both Sr {\sc ii} and Y {\sc ii} computed at electron temperatures in the range 1.0E+03 $\leq$ T(K) $\leq$ 1.0E+05. The corresponding data for all other transitions considered are available as supplementary data for each ion. In addition, for those modellers who prefer non-Mawellian averaged collision strengths, the source collision cross section data are also available from the authors on request.

\section{TARDIS 1D LTE Radiative Transfer Modelling}

In this section, we present a differential comparison of synthetic spectra to quantify the impact of our new atomic data calculations on KNe spectra. We use the Monte Carlo radiative transfer code \textsc{tardis} \cite{KerzendorfSimTardis,kerzendorf_2023_8244935} and compare to the work of \cite{gillanders2022modelling}, hereafter abbreviated to G22. This approach allows us to quantify the impact of the atomic data in models that contain a realistic treatment of the radiation transport and that provide a good match to the X-Shooter observations of AT2017gfo by \cite{Pian17} and \cite{Smartt2017}. We focus on two epochs: 1.4 and 4.4 days post-merger. These were selected as the photospheric approximation of a blackbody (as adopted by {\sc tardis}) remains well motivated at these phases  \citep{gillanders2024modelling} while also allowing investigation of how the impact of the atomic data changes in time.

The models are constructed based on composition profiles adapted by G22 from theoretical r-process nucleosynthesis calculations for binary neutron star mergers \citep{goriely_2011, bauswein_2013}. We adopt the same set of model parameters found by G22 to provide the best match to X-Shooter observations of AT2017gfo at our chosen epochs - 1.4 and 4.4 days post merger - to provide a true comparison. Full details are contained within that work, however, for completeness we include a summary here. 
Simulations are performed using \textsc{tardis} v2024.01.08, using the full relativistic treatment developed by \cite{vogl2019} due to the high ejecta speeds associated with kilonovae. In keeping with our aim to compare to the work of G22, we use the same \textsc{tardis} settings: the dilute-LTE approximation for excitation, LTE for ionisation, and \textsc{macroatom} for line interactions. 
We adopt a power law density profile for our models, of the form
\begin{equation}
    \rho (v,t_{\text{exp}}) = \rho_0 \left(\frac{t_0}{t_{\text{exp}}}\right)^3 \left(\frac{v}{v_0}\right)^{-\Gamma}
    \label{eqn:density_power_law}
\end{equation}
for ejecta velocities $v_{\text{min}}\leq v \leq v_{\text{max}}$ with constant values for $\rho_0$, $t_0$, $v_0$, $\Gamma$ and $v_{\text{max}}$. For all models, the values of $t_0 = 2$ days, $v_0 = 140000$ km s$^{-1}$, $v_{\text{max}} = 0.35$ c and $\Gamma = 3$ are used in keeping with the fitting performed by G22 previously. Other parameters - $t_{\text{exp}}$, continuum temperature, $v_{\text{min}}$, and the normalisation value for the density profile $\rho_0$ - are used as shown in \autoref{tab:tardis_params}.
Throughout our analysis we change only the atomic data input into the simulations and the elemental composition of the ejecta. We construct a baseline atomic dataset from the same sources detailed by G22 and an additional atomic dataset where we exchange the data for Sr {\sc ii} and Y {\sc ii} - previously taken from the extended Kurucz data \cite{Kurucz1995} - for the data presented in this paper. We compare these two datasets across different compositions. Initially, for illustrative purposes, we present a composition comprised purely of the elements of interest, followed by an exploration of the composition favoured by G22. 

\begin{table*}
\centering
    \begin{tabular}{cccc}
    \toprule
    \multicolumn{4}{c}{Model Parameters}       \\ \midrule
    Epoch (days) & Continuum Temperature (K) & $v_{min}$ (c) & $\rho_0$ (E-15 g cm $^{-3}$)                \\
    1.4     & 4500    & 0.28    & 12.0                  \\
    4.4     & 3200    & 0.12    & 4.0                    \\ \bottomrule
    \end{tabular}
    \caption{Parameters used for all {\sc TARDIS} models, selected to match the best fit models as explored by G22.}
    \label{tab:tardis_params}
    \end{table*}

We initially construct two kilonova models rich in the elements of interest: in one a pure Sr composition and pure Y in the other. Whilst this is not a physically realistic composition, it allows for a clean differential comparison on the effects of changing the atomic data for these species and illustrates where the spectra are most sensitive to changes in the atomic dataset. We hold all other parameters consistent for simulations of both epochs as outlined above.

\begin{figure*}
    \centering
    \includegraphics[width=1\linewidth]{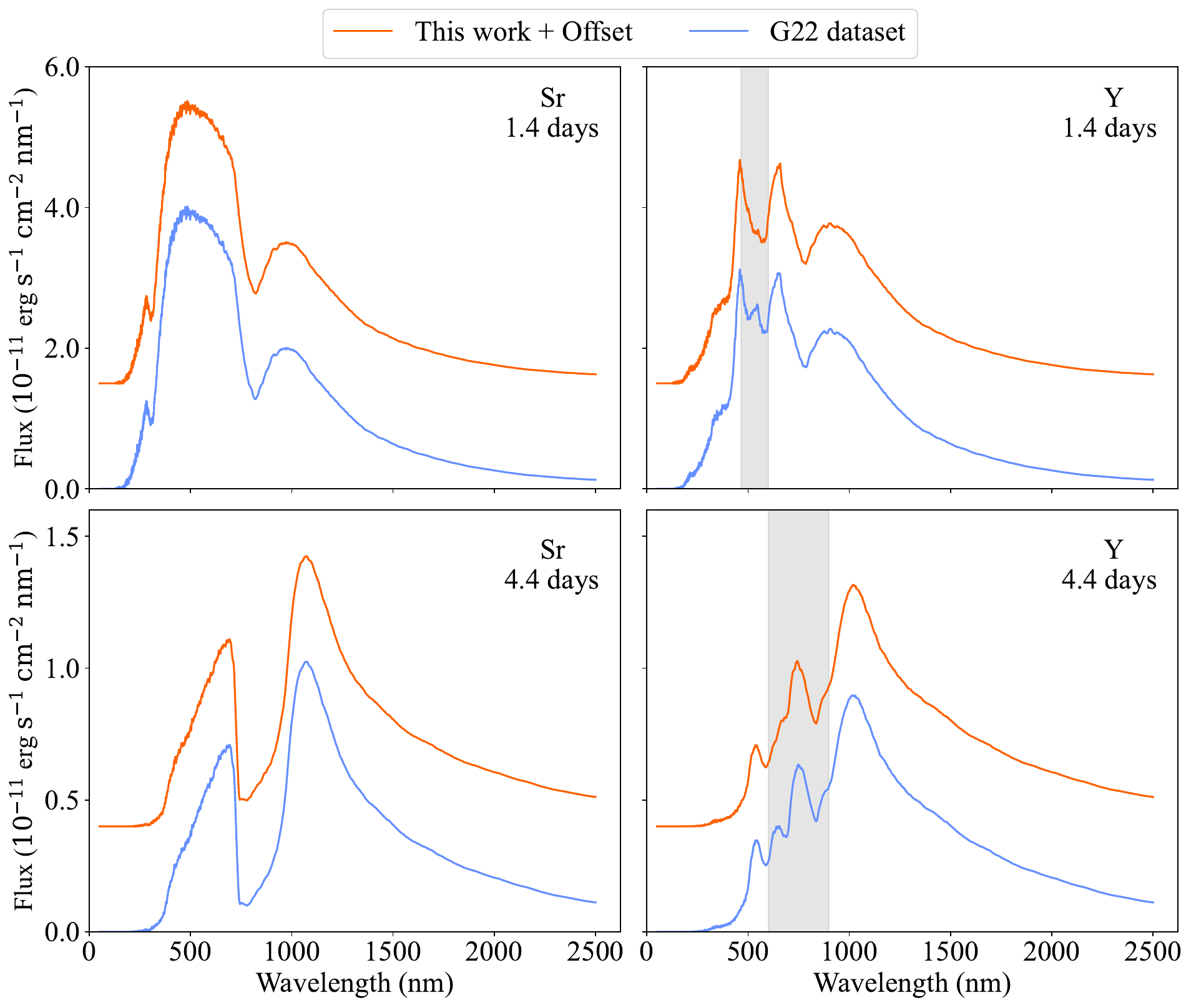}
    \caption{Two \textsc{tardis} models depicting the effects of changing the atomic data assuming a merger composed of pure Sr and pure Y. All other model parameters are selected to match the work of G22. The spectra have been arbitrarily offset for visual clarity, by 1.5E+-11 erg s$^{-1}$ cm$^{-2}$ nm$^{-1}$ for the 1.4 day epoch and 4.0E-12 erg s$^{-1}$ cm$^{-2}$ nm$^{-1}$ for the 4.4 day time frame. Grey shaded bars indicate wavelength ranges of particular interest.}
    \label{fig:pureSrYcomp}
\end{figure*}

\autoref{fig:pureSrYcomp} shows the sensitivity of the spectra to the changes in the atomic data for our pure Sr and Y models. The figure illustrates the relative insensitivity of the synthetic spectra to changes to the Sr data, as calculations with the different atomic datasets for pure Sr models are extremely similar, for both epochs. In contrast, the new Y data result in clear changes to the spectra for the pure-Y model at both epochs. The effect is most evident in the grey shaded region of around 550 nm in the 1.4 day epoch (top right panel), likely due to changes in the strengths of the 4d$^2$ $^3$P$_{2}$ $\to$ 4d5p $^3$P$^{\mathrm{o}}_{2}$ ($11\to28, \lambda = 549.89$ nm), 4d$^2$ $^3$F$_{2}$ $\to$ 4d5p $^1$D$^{\mathrm{o}}_{2}$ ($6\to18, \lambda = 551.14$ nm), and 4d$^2$ $^1$G$_{4}$ $\to$ 4d5p $^1$F$^{\mathrm{o}}_{3}$ ($13\to29, \lambda = 566.45$ nm) transitions, causing them to blend and saturate the absorption in this spectral region. Considering the bottom right panel of \autoref{fig:pureSrYcomp}, and the 4.4 day epoch, a similar conclusion can be drawn for the difference in the spectra in the shaded 670 nm region: likely caused by the 4d$^2$ $^3$P$_{2}$ $\to$ 4d5p $^3$D$^{\mathrm{o}}_{3}$ ($11\to25, \lambda = 661.56$ nm) and 4d$^2$ $^3$P$_{1}$ $\to$ 4d5p $^3$D$^{\mathrm{o}}_{2}$ ($10\to24, \lambda = 679.73$ nm) transitions.

To understand the changes in the synthetic spectra comprised of pure Y we compare both the A-values and the rates specific transitions occur in this wavelength region for both datasets. For transitions included in both this calculation, and the original Kurucz data, we compare the A-values and find an average percentage change of 85.8\%. However, the transitions highlighted above show a much smaller than average change in A-value: for example, the 566.45 nm transition occurs with a high rate at the 1.4 epochs in our model, but shows a percentage increase in A-value of approximately 1\% (\autoref{tab:yii_lines}). We additionally compare the total number of interactions this line is responsible for in our models. Here we see an increase of 2.6\% from the Kurucz dataset to our calculation. A similar pattern is seen for all the lines highlighted above. We therefore conclude that although these transitions are responsible for the most interactions in our models, the consistency of their A-values across datasets means they are unlikely to be responsible for a visible change to our synthetic spectra. We instead look to other transitions in the surrounding wavelength range with a more dramatic change in number of transitions to explain the changes in spectral shape. In the region around 500 nm, we find two transitions with a large increase in the number of interactions: the 4d$^2$ $^1$S$_{0}$ $\to$ 5s5p $^1$P$_{1}$ ($17\to30, \lambda = 512.86$ nm) and 4d$^2$ $^3$D$_{4}$ $\to$ 4d5p $^3$D$_{3}$ ($8\to21, \lambda = 532.23$ nm) transitions, with a 77.3 and 167.0\% increase respectively. This is echoed by a change in A-value for these lines, with an increase in this calculation against Kurucz of 92.7 and 263.6\%. From this we conclude that, although there is little change to the A-values and therefore number of interactions occurring for the strongest lines in our models, a substantial increase in interactions with multiple surrounding weaker lines is likely responsible for the changes visible in our synthetic spectra. 

\begin{figure*}
    \centering
    \includegraphics[width=1\linewidth]{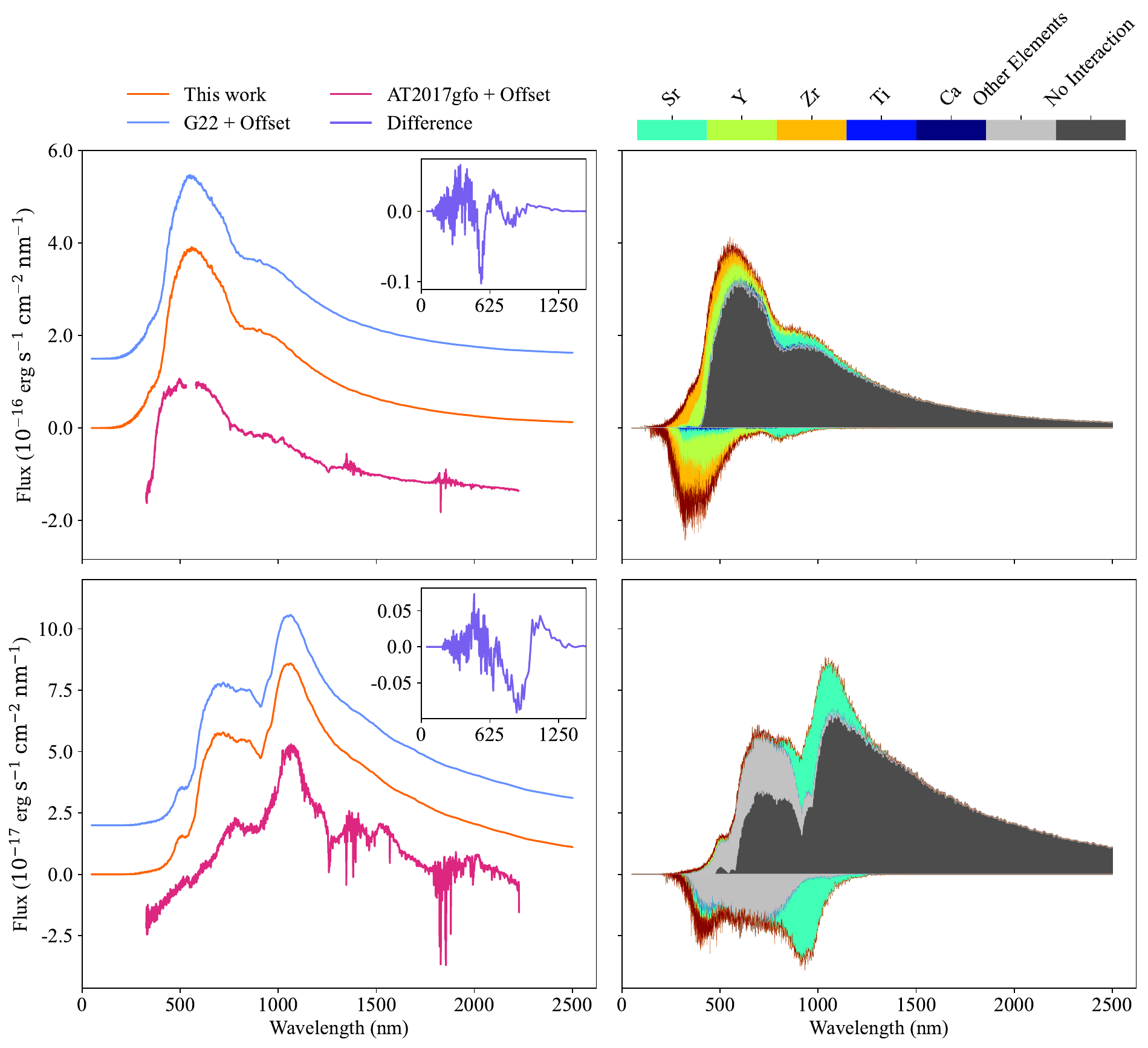}
    \caption{\textsc{tardis} models of the full spectrum of AT2017gfo at 1.4 (top row) and 4.4 (bottom row) days post-merger. The left column displays the shape of the spectra in comparison to observations of AT2017gfo published by \citet{Pian17} and \citet{Smartt2017}, with inset plots of the difference between using the G22 dataset and the new calculation presented here. The observed spectra and that from the G22 dataset have been arbitrarily offset from the new calculation for visual clarity, by $\mp$ 1.5E-16 erg s$^{-1}$ cm$^{-2}$ nm$^{-1}$ at 1.4 days and $\mp$ 2.0E-17 erg s$^{-1}$ cm$^{-2}$ nm$^{-1}$ at 4.4 days. The right column shows the elemental composition of the spectra.}
    \label{fig:SDEC}
\end{figure*}

We construct a model using a dataset built from the same sources as outlined by G22, based on solar r-process abundances to serve as a baseline for our comparisons. To show the effect of our new calculations on the spectra, we create one further dataset using the same sources for all other elements but replacing the previously used Sr and Y data with that calculated in this work. Our atomic calculations and model are consistent with the findings of G22, however we note that our modelled flux is higher than that of the observations of AT2017gfo and that of the synthetic spectra shown by G22. This is due to recent improvements in the special relativistic treatment in \textsc{tardis} \citep{vogl2019}. Although the normalisation differs, the spectral features remain consistent and our parameters are identical to those used by G22. For consistency, we run both calculations with the latest code version - including the updated relativistic treatment - enabling us to present a true differential comparison.  Comparing the spectra presented in the top left panel of \autoref{fig:SDEC}, to observations from the 1.4 day epoch we see only a subtle effect - most notable in the region surrounding the 810 nm Sr {\sc ii} feature identified by \cite{Watson19} and in the region of peak flux - caused by the changed atomic data. This is most clearly visible in the inset difference plot. The peak of the blackbody becomes marginally rounder and broader, while the 810 nm feature becomes deeper. The sensitivity of this feature to the change in atomic data presented here is in agreement with the corresponding identification of the 4d $\to$ 5p transition array of Sr {\sc ii} from \cite{Watson19}.

For the G22 model at 4.4 days, spectra obtained with our new atomic data are almost identical (within the Monte Carlo noise) to those obtained with the G22 atomic dataset, with only very small changes visible in the difference plot. This can be understood since the contribution of Y in this model at this phase is minor (see G22 and \autoref{fig:SDEC}) and, as noted above, the differences in the atomic data do not dramatically alter the Sr \textsc{ii} feature. This is potentially due to the evolving conditions in the expanding ejecta which changes the strength of features. In particular, as shown by G22 for this particular model, changes in density and ionisation mean that the contribution of Y decreases significantly while other elements start to have a larger impact: such as the dominance of line blanketing from lanthanide species (for discussion see G22). 

Although \autoref{fig:SDEC} shows only small changes to the full spectrum with the refinement of the atomic data set used in \textsc{tardis} modelling when compared to the work of G22, the pure Sr and pure Y compositions in \autoref{fig:pureSrYcomp} highlight the potential sensitivity to individual transitions. As such, refinements of the atomic data available will become of even higher importance as models develop further, particularly with the aim of working in NLTE. Although generating full synthetic spectra under NLTE conditions is beyond the scope of this work, we find there will be non-negligible differences if switching to a collisionally dominated domain, as investigated below.

\section{ ColRadPy NLTE Collisional Radiative Modelling}

Previous works have shown that, as the NSM ejecta expands with increasing time and decreasing density and temperature, the KNe transitions from LTE to NLTE. For example \cite{Pognan2023} predicted that even a few days post merger the differences between LTE and NLTE opacities could be as large as several orders of magnitude for some r-process ion stages. In order to investigate the spectral evolution of the KNe AT2017gfo, we present in this section a NLTE analysis utilising the Generalized Collisional Radiative theory (GCR) of \cite{Summers2006} which was based on the Collisional Radiative theory of \cite{bates1962recombination}. This theoretical modelling approach has been scripted into a Python package, {\sc ColRadPy}, by \cite{johnson2019colradpy}. This modelling package allows the energy levels, A-values, effective collision strengths and most significantly the excitation rate coefficients produced in the Sr {\sc ii} and Y {\sc ii} calculations outlined above, to be employed in the modelling of a wide range of plasma parameters to include electron temperature and electron density. In this paper, we will limit our modelling to conditions similar to those of the electromagnetic counterpart of GW170817, AT2017gfo. The modelling code {\sc ColRadPy} embodies the solution of the full collisional-radiative equations,
\begin{equation}
    \frac{\dd N_j}{\dd t} = \sum_{i} C_{ij} N_i,
\end{equation}
where $C_{ij}$ is the collisional-radiative matrix accounting for radiative and collisional excitations/de-excitations which are employed in this work. It can additionally include ionisation and recombination rates we do not consider here. For astrophysical plasmas it is adequate to quote the steady state, or quasi-static, solution.

\begin{figure}
    \centering
    \includegraphics[width=1\linewidth]{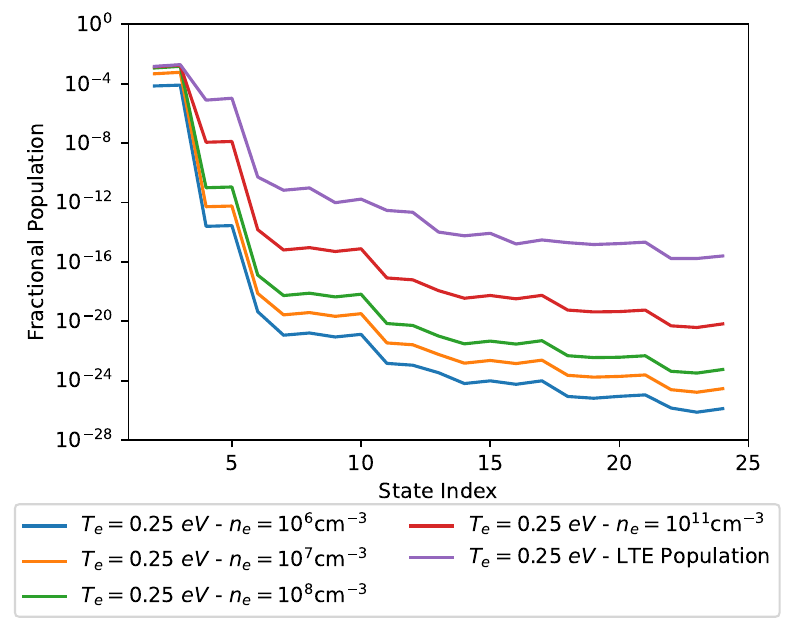}
    \caption{Sr {\sc ii} level populations at $T_e$ = 0.25 eV $\approx $ 2900K and a range of electron densities. The LTE populations are also shown.}
    \label{fig:srii_pop}
\end{figure}

\begin{figure}
    \centering
    \includegraphics[width=1\linewidth]{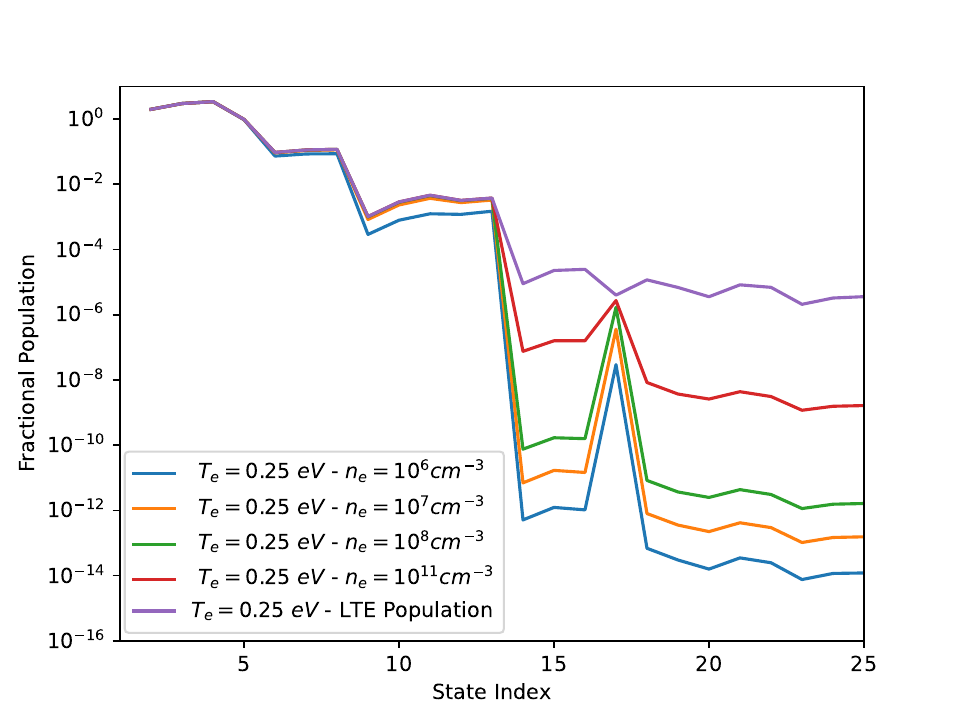}
    \caption{Y {\sc ii} level populations at $T_e$ = 0.25 eV $\approx $ 2900 K and a range of electron densities. The LTE populations are also shown. }
    \label{fig:yii_pop}
\end{figure}

To investigate the level populations we present in Figures \ref{fig:srii_pop} and \ref{fig:yii_pop} the fractional Sr {\sc ii} and Y {\sc ii} populations for the lowest 25 states of each ion at electron temperatures of $T_e$ = 0.25 eV and a range of electron densities 10$^6 \leq  n_e \leq $ 10$^{11}$ cm$^{-3}$. This temperature was chosen based on the analysis of \cite{gillanders2024modelling} for Sr {\sc ii} and \cite{SnepWat23} for Y {\sc ii}. Also shown in these figures are the corresponding LTE populations for the same temperature values. Here, LTE is reached in the limit of large densities. LTE can also be attained in the radiation-dominated phase in the early times of the KNe, which is not accounted for in this model. The analysis here is primarily concerned with the later collisionally dominated epochs. Clearly, the departure from LTE populations occurs at a relatively low index in the case of Sr {\sc ii} due to the fact that the first dipole E1 transition occurs at index number 4.  For Y {\sc ii}, on the other hand, LTE populations largely hold for the lowest 14 target levels after which deviations from LTE appear significant when the first dipole line appears at level index 14. However, it is noteworthy that these deviations from LTE occur at around the same excitation energy of $\sim 0.2$ Ry. At these level indices, it is clear that the density and collision rates are unable to outweigh the Einstein A-coefficients, and LTE is unattainable for these levels.

\begin{figure}
    \centering
    \includegraphics[width=\linewidth]{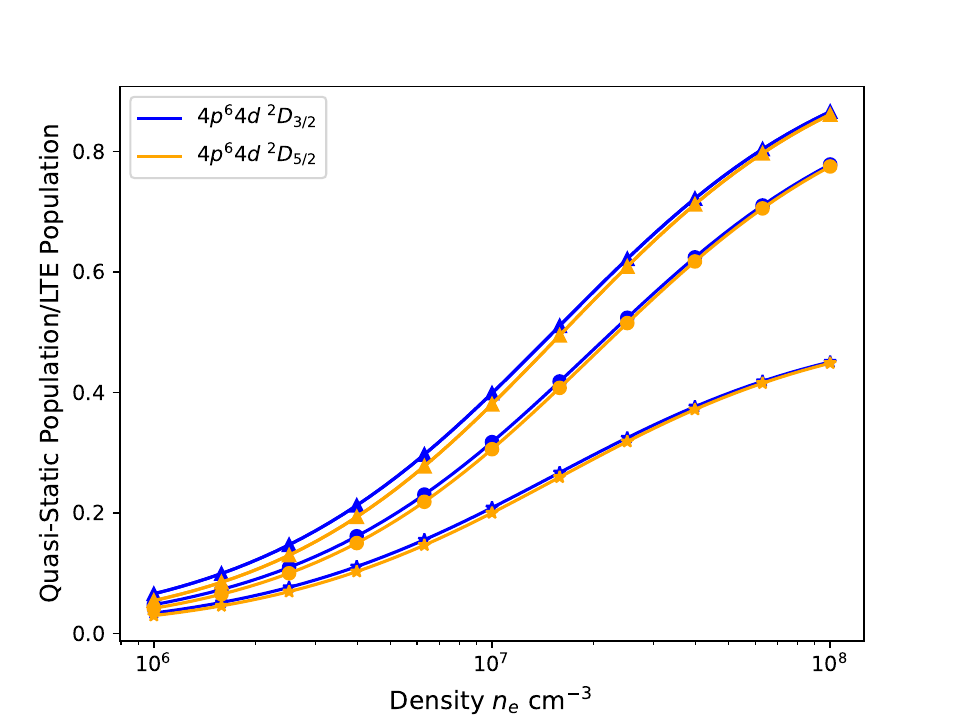}
    \caption{Ratio of populations within a quasi-static approximation and LTE as a function of electron density (cm$^{-3}$) for the first 2 excited states of Sr {\sc ii}. The symbols on the plot represent electron temperatures: $\Delta$ - 0.1 eV, $\bigcirc$ - 0.25 eV, $\bigstar$ - 0.55 eV respectively. }
    \label{fig:sr_ii_nlte}
\end{figure}

\begin{figure}
    \centering
    \includegraphics[width=\linewidth]{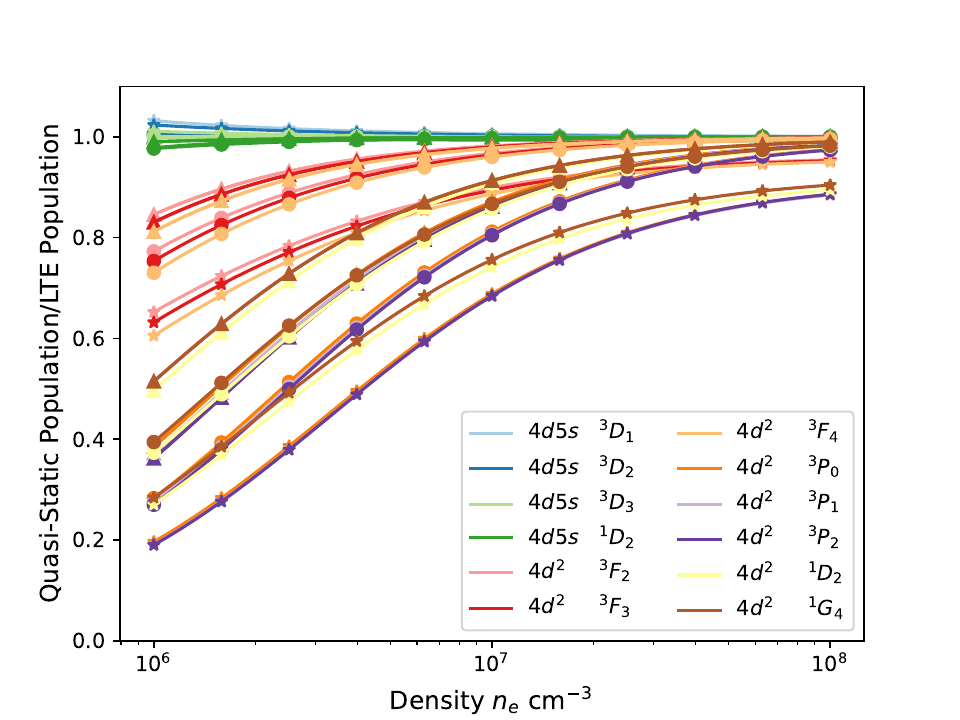}
    \caption{Ratio of quasi-static populations and LTE populations as a function of electron density (cm$^{-3}$) for the first 12 Y {\sc ii} excited states. The symbols on the plot represent electron temperatures: $\Delta$ - 0.1 eV, $\bigcirc$ - 0.25 eV, $\bigstar$ - 0.55 eV respectively.}
    \label{fig:y_ii_nlte}
\end{figure}

To investigate further the deviation from LTE conditions, we present in Figures \ref{fig:sr_ii_nlte} and \ref{fig:y_ii_nlte} the ratio of the steady state (NLTE) and the LTE populations as a function of electron density in cm$^{-3}$  for the low lying levels of Sr {\sc ii} and Y {\sc ii}. Here we investigate a similar parameter space to that of \cite{gillanders2024modelling}, who calculated $n_e$ in the regime of $10^6 - 10^8$ cm$^{-3}$ with $T_e$ in the range of 0.25 - 0.4 eV (2500 - 4500 K. In this work we consider a slightly larger temperature range with $T_e$ = 0.10, 0.25 and 0.55 eV (1100, 2900 and 6400 K). For the case of Sr {\sc ii}, both the 4p$^6$4d $^{2}$D$_{3/2}$ and 4p$^6$4d $^{2}$D$_{5/2}$ states exhibit significant departure from LTE populations, by up to a factor of 10 for the lowest densities considered (10$^{6}$ cm$^{-3}$) and across all temperatures. For the 12 states considered for Y {\sc ii} it is clear from Figure \ref{fig:y_ii_nlte} that those with configurations 4d5s show no deviation from LTE across the temperature and density ranges considered. In contrast the higher lying levels with configuration 4d$^{2}$ show deviations from LTE by nearly a factor of 5. We can therefore conclude, and support the NLTE modelling of Au {\sc i} by \cite{McCann2022},
that the metastable level populations can deviate substantially from their LTE proportions at conditions relevant to NSM events. In addition, excitation from
these radiatively metastable levels can dominate direct excitation from the
ground state.


In the remainder of this section we investigate the excitation photon emissivity coefficient (PEC), often useful for predicting individual spectral line emission. A PEC is a derived coefficient that is associated with a single spectral line and its excitation component (the full PEC is a combination of excitation, recombination and change exchange components) is given by 
\begin{equation}
    \mathrm{PEC}_{j \rightarrow i}^{\mathrm{excit}} = \frac{N_j^{\mathrm{excit}} A_{j \rightarrow i}}{n_e}
\end{equation} 
where $N_j^{\mathrm{excit}}$ is the weighted population of the upper level  $j$ defined so that
\begin{equation}
N_{j}^{\mathrm{excit}} = \frac{N_j}{N_1}
\end{equation}
where $N_j$ is the population of the upper level and $N_1$ is the population of the ground state. As defined previously $A_{j \rightarrow i}$ is the Einstein A-coefficient for the transition from $j$ to $i$ and $n_e$ is the electron density in cm$^{-3}$.

\begin{figure*}
    \centering
    \includegraphics[width=1\linewidth]{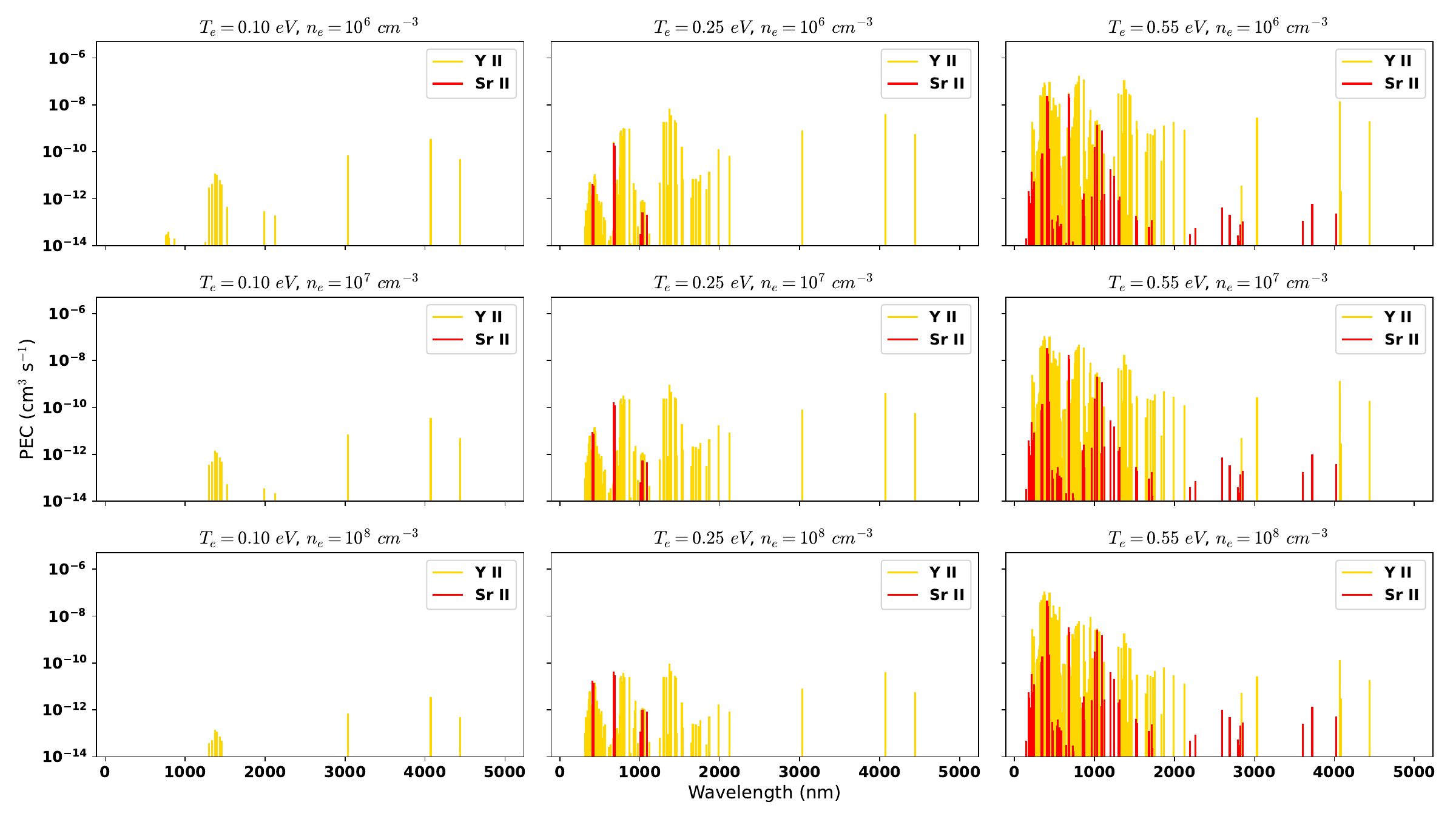}
    \caption{The Photon Emmisivity Coefficients (PECs) are shown for transitions with wavelength between 0 and 5000nm for single element plasmas of both Sr II and Y II at 3 different temperatures and 3 different densities, provided the PEC $ > 10^{-14} $ (cm$^{3}$ s$^{-1}$) }
    \label{fig:comb_facetplot}
\end{figure*}

\begin{figure}
    \centering
    \includegraphics[width=0.9\linewidth]{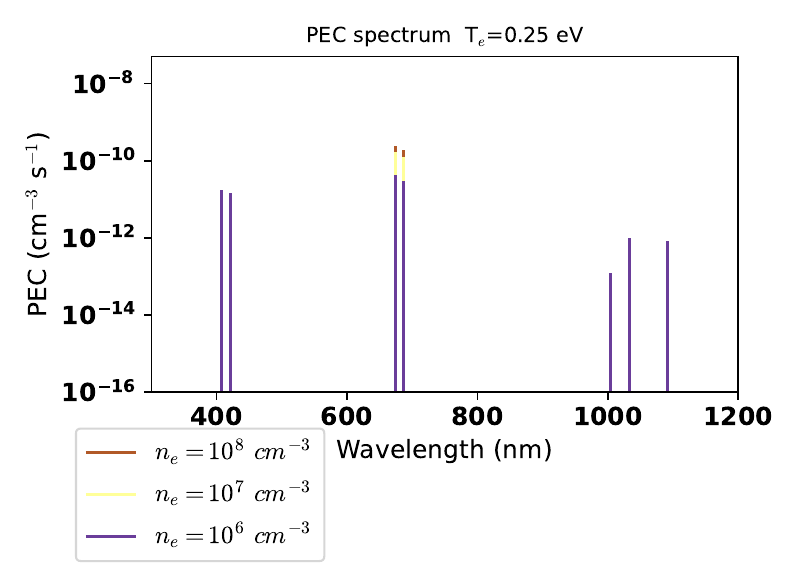}
    \caption{The excitation Photon Emissivity Coefficients (PEC) for the 300-1200 nm wavelegth window in both temperature and density space for Sr {\sc ii}.}
    \label{fig:Sr_II_1000n}
\end{figure}

\begin{figure}
    \centering
    \includegraphics[width=0.9\linewidth]{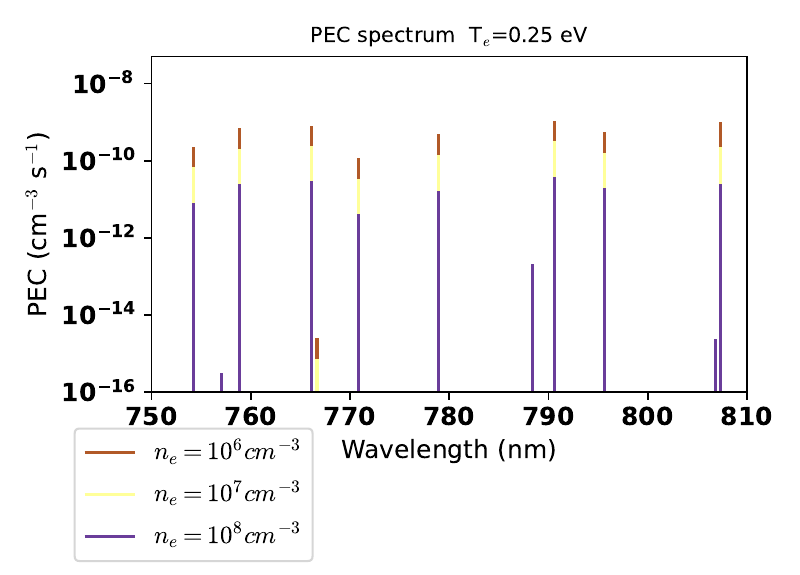}
    \caption{The excitation Photon Emissivity Coefficients (PEC) for the 750-810 nm wavelength window of Y {\sc ii} transitions in both temperature and density space. }
    \label{fig:yii_780}
\end{figure}

\begin{figure}
    \centering
    \includegraphics[width=0.9\linewidth]{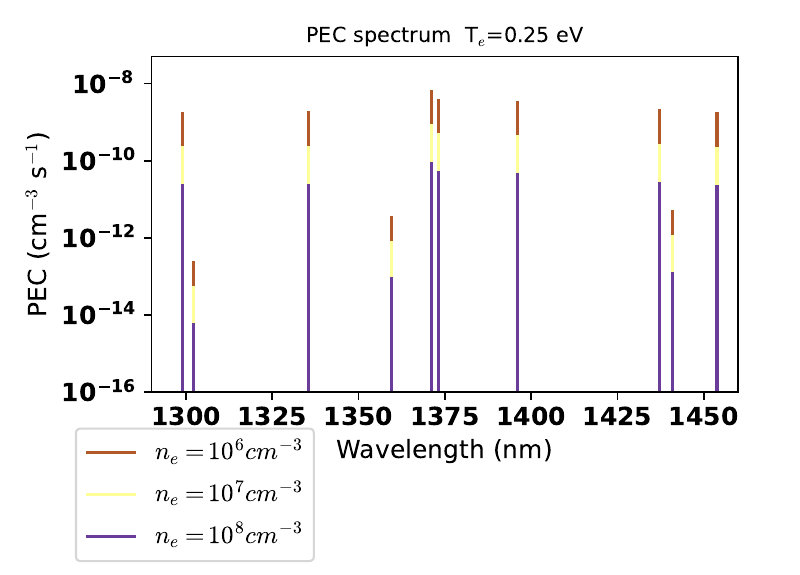}
    \caption{The excitation Photon Emissivity Coefficients (PEC) for the 1290-1460 nm wavelength window of Y {\sc ii} transitions in both temperature and density space.}
    \label{fig:yii_1290}
\end{figure}

In Figure \ref{fig:comb_facetplot} we plot these PEC coefficients (in units cm$^3$ s$^{-1}$) as a function of wavelength (nm) from 0 to 5000 nm spanning the spectrum from the UV to the IR. Three electron temperatures are considered in the computations, $T_e$ = 0.1, 0.25 and 0.55 eV, and electron densities $n_e$ = 10$^6$, 10$^7$ and 10$^8$ cm$^{-3}$ of relevance to KNe modelling. Clearly there is significant line emission across the spectrum for both elements at the temperatures and densities of interest.  Particularly clear are three Sr {\sc ii} emission lines between 1000 and 1100nm, the 1003.94nm line corresponding to transition 2-5 , the 1033.01nm line for transition 3-5  and the 1091.79nm line for transition 2-4. This region of the spectrum corresponds to the wavelength region where the well identified P Cygni spectral line of Sr {\sc ii} was predicted by \cite{Watson19} at
approximately 1$\mu$m. Figure \ref{fig:Sr_II_1000n} focuses on the wavelength band 350 - 1200 nm at 0.25 eV where it is seen that while these three Sr {\sc ii} are reasonably strong, in conformity with \cite{gillanders2024modelling}, we also predict strong emission at  $\sim 400$ and $650$ nm. These are due to the 1-5, 1-4, 1-3 and 1-2 lines at 407.89, 421.67, 674.03 and 687.01nm respectively. As discussed in the cited article, the prominence of the resonance lines at $\sim 650$ nm contrasts with the lack of strong features in observations and potentially places constraints on the mass of Sr {\sc ii} in the kilonova, as well as the geometry of the ejecta \citep{gillanders2024modelling,collins24}. 

Attempts were also made to probe the emission of the 788nm line to which \cite{SnepWat23} attributed the 760nm Y {\sc ii} P Cygni feature in AT2017gfo. This wavelength region exhibits dense line emission in Figure \ref{fig:comb_facetplot} so in an attempt to magnify this region we plot in Figure \ref{fig:yii_780} the PEC plot from 750 - 810 nm and again the temperature used in the simulations was $T_e$ = 0.25 eV to follow the modelling of \cite{SnepWat23}. It is evident that the strongest emission in NLTE modelling is not the 788.41nm line (transition index 12-20) as there are several stronger lines in this wavelength window, the strongest of which are the 790.62nm (transition index 4-11) and the 807.32nm (transition 5-13). 
 It is only at a density of $\sim$ 10$^{11}$cm$^{-3}$ that the fractional population is high enough for the transition indexed 12-20 to be the strongest line in this wavelength window, a density which is unlikely in a KNe event such as AT2017gfo. In high density LTE modelling, however, the fractional population in comparison to ground is several orders of magnitude stronger, and emission would then be consistent with the findings of \cite{SnepWat23}.

A recent publication by \cite{gillanders2024modelling} when modelling the spectra of AT2017gfo, searched for potential candidate ions prominent in the spectra. Emission features in the late phase spectra were identified at $\sim$ 0.79, 1.08, 1.23, 1.40, 1.58, 2.059 and 2.135$\mu$m.  The intensities were computed with $T = 0.4$ eV and the computations were based on data from \cite{Kurucz1995}. 
In Table \ref{tab:yii_lines_jg} we list all the Sr {\sc ii} and Y {\sc ii} transitions which emerged from their line analysis to include the wavelengths, transition index, lower and upper level terms and energies. A comparison was made between the present {\sc grasp$^0$} A-values and \cite{Kurucz1995} and excellent agreement was found for all lines considered. The PEC plot covering the wavelength region 750-810nm has already been presented in Figure \ref{fig:yii_780} and the lines clearly match the predictions made by \cite{gillanders2024modelling} for all three densities considered. To investigate the wavelength region for the remaining lines we plot in Figure \ref{fig:yii_1290}
the PEC coefficients for wavelengths between 1290 and 1460nm, computed at the same temperature $T$ = 0.25 eV, and clearly evident is the presence of the remaining 7 lines of interest, again at all three densities considered.  
\begin{table*}
    \centering
        \begin{tabular}{ccrcrrccc}
            \hline
            \vspace{2mm}
           
            $\lambda$ &Index & $E_{{i}}$\phantom{00} & Lower & $E_{{j}}$\phantom{00} & Upper\phantom{-} & 
            \multicolumn{2}{c}{$A_{{j\to i}}$ (s$^{-1}$)} & $L$ \\  
            (nm)&(${i}$-${j}$)\phantom{0} & (cm$^{-1}$) & ${i}$ &   (cm$^{-1}$) &  ${j}\phantom{00}$ & Present & Kur95 & ($10^{50}\frac{M_{ion}}{10^{-3}M_\odot}$ ph s$^{-1}$ ) \\ 
            \vspace{-3mm}\\\hline
            \\
            Sr {\sc ii} lines &&&&&& \vspace{1mm}
            \\
            \emph{Allowed} \\
            407.89 &  1 - 5   &       0.00 & 5s $^2$S$_{1/2}$ &  24516.65 & 5p $^2$P$^{\mathrm{o}}_{3/2}$ & 1.58E+08 & 1.41E+08 & 5.93E-04 \\
            421.67 &  1 - 4   &       0.00 & 5s $^2$S$_{1/2}$ &  23715.19 & 5p $^2$P$^{\mathrm{o}}_{1/2}$ & 1.44E+08 & 1.26E+08 & 4.81E-04 \\
           1003.94 &  2 - 5   &   14555.90 & 4d $^2$D$_{3/2}$ &  24516.65 & 5p $^2$P$^{\mathrm{o}}_{3/2}$ & 1.10E+06 & 9.97E+05 & 4.13E-06 \\
           1033.01 &  3 - 5   &   14836.24 & 4d $^2$D$_{5/2}$ &  24516.65 & 5p $^2$P$^{\mathrm{o}}_{3/2}$ & 9.22E+06 & 8.79E+06 & 3.46E-05 \\
           1091.79 &  2 - 4   &   14555.90 & 4d $^2$D$_{3/2}$ &  23715.19 & 5p $^2$P$^{\mathrm{o}}_{1/2}$ & 8.61E+06 & 7.46E+06 & 2.88E-05 \\ 
           \emph{Forbidden} \\
           674.03 &  1 - 3   &       0.00 & 5s $^2$S$_{1/2}$ &  14836.24 & 4d $^2$D$^{ }_{5/2}$ & 2.91E+00 & 2.56E+00 & 3.15E-02 \\
           687.01 &  1 - 2   &       0.00 & 5s $^2$S$_{1/2}$ &  14555.90 & 4d $^2$D$^{ }_{3/2}$ & 2.62E+00 & 2.30E+00 & 2.48E-02 \\
           \\
            Y {\sc ii} lines &&&&&& \vspace{1mm}
            \\
            \emph{Allowed}\\
            661.56 &  11 - 25 & 14098.07 &  4d$^2$ $^3$P$_{2}$ &  29213.96 & 4d5p $^3$D$^{\mathrm{o}}_3$ & 3.72E+06 & 1.69E+06 & 5.69E-07 \\                         
            679.72 &   9 - 23 & 13883.38 &  4d$^2$ $^3$P$_{0}$ &  28595.28 & 4d5p $^3$D$^{\mathrm{o}}_1$ & 1.15E+06 & 1.74E+06 & 1.11E-07 \\                         
            679.73 &  10 - 24 & 14018.27 &  4d$^2$ $^3$P$_{1}$ &  28730.00 & 4d5p $^3$D$^{\mathrm{o}}_2$ & 2.73E+06 & 2.51E+06 & 4.03E-07 \\                         
            683.44 &  11 - 24 & 14098.07 &  4d$^2$ $^3$P$_{2}$ &  28730.00 & 4d5p $^3$D$^{\mathrm{o}}_2$ & 7.19E+05 & 3.28E+05 & 1.06E-07 \\                         
            686.01 &  10 - 23 & 14018.27 &  4d$^2$ $^3$P$_{1}$ &  28595.28 & 4d5p $^3$D$^{\mathrm{o}}_1$ & 7.97E+05 & 9.02E+05 & 7.66E-08 \\                         
            689.79 &  11 - 23 & 14098.07 &  4d$^2$ $^3$P$_{2}$ &  28595.28 & 4d5p $^3$D$^{\mathrm{o}}_1$ & 3.20E+05 & 1.52E+05 & 3.08E-08 \\                         
            726.62 &  12 - 23 & 14832.86 &  4d$^2$ $^1$D$_{2}$ &  28595.28 & 4d5p $^3$D$^{\mathrm{o}}_1$ & 8.01E+06 & 1.33E+06 & 7.70E-07 \\                         
            733.50 &   9 - 20 & 13883.38 &  4d$^2$ $^3$P$_{0}$ &  27516.69 & 4d5p $^1$P$^{\mathrm{o}}_1$ & 6.66E+05 & 5.52E+04 & 1.34E-07 \\                         
            745.23 &  11 - 20 & 14098.07 &  4d$^2$ $^3$P$_{2}$ &  27516.69 & 4d5p $^1$P$^{\mathrm{o}}_1$ & 3.59E+04 & 3.63E+05 & 7.22E-09 \\                         
            788.41 &  12 - 20 & 14832.86 &  4d$^2$ $^1$D$_{2}$ &  27516.69 & 4d5p $^1$P$^{\mathrm{o}}_1$ & 8.66E+06 & 9.63E+06 & 1.74E-06 \\
            \emph{Forbidden} \\
            754.27 & 2 - 11&   840.20 & 5s4d $^3$D$_1$ &  14098.07 & 4d$^2$ $^3$P$_2$    &  1.82E-01 & 1.83E-01 & 2.87E-03 \\                       
            758.84 & 2 - 10&   840.20 & 5s4d $^3$D$_1$ &  14018.27 & 4d$^2$ $^3$P$_1$    &  8.72E-01 & 8.95E-01 & 8.74E-03 \\                       
            766.11 & 3 - 11&  1045.08 & 5s4d $^3$D$_2$ &  14098.07 & 4d$^2$ $^3$P$_2$    &  6.50E-01 & 6.71E-01 & 1.02E-02 \\                       
            770.82 & 3 - 10&  1045.08 & 5s4d $^3$D$_2$ &  14018.27 & 4d$^2$ $^3$P$_1$    &  1.45E-01 & 1.51E-01 & 1.45E-03 \\                       
            778.92 & 3 - l9&  1045.08 & 5s4d $^3$D$_2$ &  13883.38 & 4d$^2$ $^3$P$_0$    &  1.66E+00 & 1.72E+00 & 6.11E-03 \\                       
            790.62 & 4 - 11&  1449.75 & 5s4d $^3$D$_3$ &  14098.07 & 4d$^2$ $^3$P$_2$    &  8.54E-01 & 8.77E-01 & 1.35E-02 \\                       
            795.64 & 4 - 10&  1449.75 & 5s4d $^3$D$_3$ &  14018.27 & 4d$^2$ $^3$P$_1$    &  6.94E-01 & 7.30E-01 & 6.96E-03 \\                       
            807.32 & 5 - 13&  3296.18 & 5s4d $^1$D$_2$ &  15682.90 & 4d$^2$ $^1$G$_4$    &  6.73E-01 & 7.41E-01 & 1.27E-02 \\                       
           1299.00 & 3 -  8&  1045.08 & 5s4d $^3$D$_2$ &   8743.32 & 4d$^2$ $^3$F$_4$    &  2.13E-02 & 2.23E-02 & 2.33E-02 \\                       
           1335.50 & 2 -  7&   840.20 & 5s4d $^3$D$_1$ &   8328.04 & 4d$^2$ $^3$F$_3$    &  2.23E-02 & 2.35E-02 & 2.40E-02 \\                       
           1371.07 & 4 -  8&  1449.75 & 5s4d $^3$D$_3$ &   8743.32 & 4d$^2$ $^3$F$_4$    &  8.09E-02 & 8.59E-02 & 8.84E-02 \\                       
           1373.07 & 3 -  7&  1045.08 & 5s4d $^3$D$_2$ &   8328.04 & 4d$^2$ $^3$F$_3$    &  4.78E-02 & 5.04E-02 & 5.15E-02 \\                       
           1396.08 & 2 -  6&   840.20 & 5s4d $^3$D$_1$ &   8003.13 & 4d$^2$ $^3$F$_2$    &  4.96E-02 & 5.28E-02 & 4.60E-02 \\                       
           1437.18 & 3 -  6&  1045.08 & 5s4d $^3$D$_2$ &   8003.13 & 4d$^2$ $^3$F$_2$    &  3.00E-02 & 3.22E-02 & 2.78E-02 \\                       
           1453.85 & 4 -  7&  1449.75 & 5s4d $^3$D$_3$ &   8328.04 & 4d$^2$ $^3$F$_3$    &  2.13E-02 & 2.31E-02 & 2.30E-02 \\
           3033.82 & 1 -  5 &    0.00 &  5s$^2$ $^1$S$_{0}$& 3296.18 & 5s4d $^1$D$_{2}$  &  8.44E-04 & 9.49E-04 & 1.02E-02 \\                        
           4071.69 & 2 -  5 &  840.20 & 5s4d $^3$D$_{1}$   & 3296.18 & 5s4d $^1$D$_{2}$  &  4.20E-03 & 5.72E-03 & 5.09E-02 \\                        
           4442.27 & 3 -  5 & 1045.08 & 5s4d $^3$D$_{2}$   & 3296.18 & 5s4d $^1$D$_{2}$  &  5.94E-04 & 8.05E-04 & 7.19E-03 \\
          
           \hline
        \end{tabular}
        \caption{The set of spectral lines highlighted for Sr {\sc ii} and Y {\sc ii}. The transition probabilities $A_{j \rightarrow i}$ calculated here are compared with those available on the NIST database and the calculations of \citep{Kurucz1995}. Additionally presented is the predicted photon luminosity for a mass of 10$^{-3}$ M$_{\odot}$ of Sr {\sc ii} and 10$^{-3}$ M$_{\odot}$ Y {\sc ii} at a temperature of 0.25 eV and electron density 10$^6$ cm$^{-3}$
         .}
        \label{tab:yii_lines_jg}
    \end{table*}
We extend this analysis by combining the calculated PECs with the respective population calculations to obtain an estimate of the photon luminosity one might expect based on the NLTE data calculated here. The photon luminosity is expressed as
\begin{equation}
    L = n_e  \frac{\text{PEC}_{i\to j}}{\sum_i N_i^{\mathrm{excit}}}  \frac{M_{\text{ion}}}{\text{m}_{\text{ion}}}, \label{eq:photon_lumo}
\end{equation}
in units of photons per time. Here $M_{\text{ion}}$ is the total mass in the ejecta of the particular ion, and $m_{\text{ion}}$ is the mass of a single ion, and their ratio in \eqref{eq:photon_lumo} simply gives the number of ions of this species in the ejecta. This quantity was calculated at a temperature of 0.25 eV and electron density of $10^6$ cm$^{-3}$and presented the final column of Table \ref{tab:yii_lines_jg}. A mass of $M_{\text{Sr}} =M_{\text{Y}} = 10^{-3}$ M$_\odot$ was employed. These parameters were chosen to reflect the calculation of \cite{gillanders2024modelling}. Clearly, the strongest emission flux emerges from the forbidden Y {\sc ii} lines highlighted in the cited article. Furthermore, we note that the photon flux estimated is comparable to the parametric fitting of the observed spectra featured in \cite{gillanders2024modelling}. In particular, they report a flux for the $1.4 \mu$m feature at +9.4 days of 8.46E+38 ergs $\text{s}^{-1}$ , corresponding to a photon flux of $\sim 6.00$E+50 $\text{s}^{-1}$, which is comparable to the  presented photon fluxes around this wavelength in Table \ref{tab:yii_lines_jg}. The contribution of lines from this table in the vicinity of this wavelength corresponds to around 2.87E+49 ph $\text{s}^{-1}$, making it plausible that Y {\sc ii} provides an appreciable contribution to this spectral feature. It is possible with further modelling with realistic ejecta compositions may provide a more conclusive explanation of this particular feature.
In further congruence  with \cite{gillanders2024modelling}, we also find that the the Sr {\sc ii } NIR triplet is significantly weaker that the forbidden  Sr {\sc ii} lines at $\sim$ 650 nm for the conditions considered (see Gillanders et al. 2024 for discussion of implications, and possible role of clumpy ejecta in affecting these line ratios).

\begin{figure*}
    \centering
    \includegraphics[width=1\linewidth]{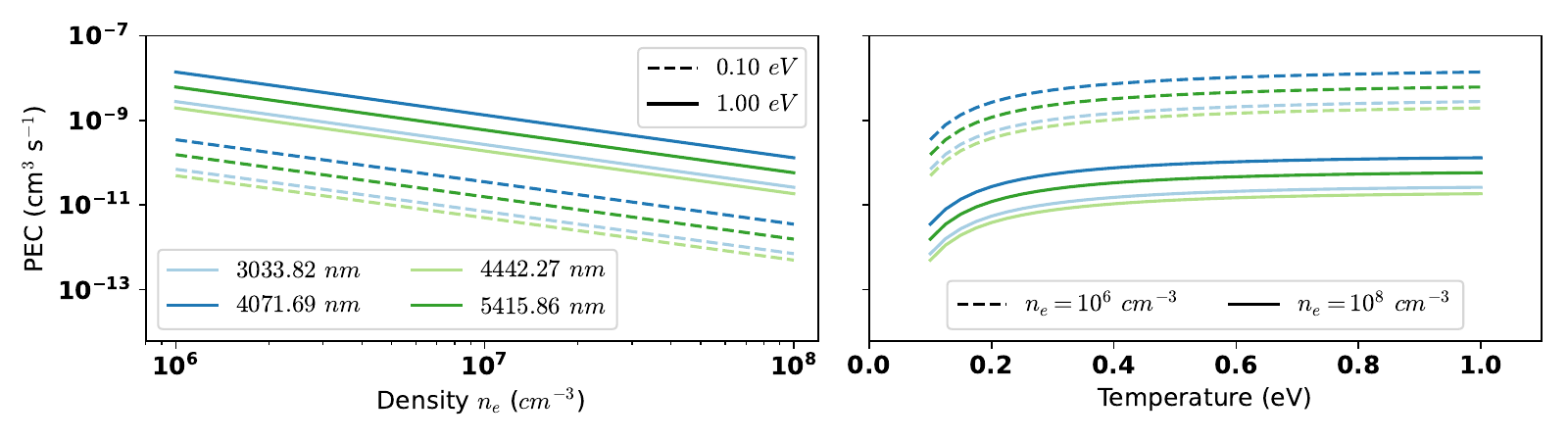}
    \caption{The excitation Photon Emissivity Coefficients (PEC) for the 5-3 (4442.27nm), 5-2 (4071.69nm) and 5-1 (3033.82nm) transitions in both temperature and density space.}
    \label{fig:5-below}
\end{figure*}

Finally, the existence or non-existence of useful temperature and density diagnostic lines across the full wavelength range from the UV to the IR is of particular interest to modellers and observers alike. In this particular context, the expectation of line blending reduces the practicality when compared with other astrophysical processes. The usefulness can be further assessed based on the above flux calculations. Nonetheless for completeness, in this section we probe potential candidates for lines of Sr {\sc ii} and Y {\sc ii} which exhibit either strong temperature or density dependence or both. We begin with Y {\sc ii} as the fourth excited state 4d5s $^1$D$_2$ proved interesting. Decay to 
the states below this level, particularly transitions 1-5, 2-5 and 3-5, are visible at 3033.82nm, 4071.69nm, 4442.27nm respectively in Figure \ref{fig:comb_facetplot}. All three of these lines are visible at all temperatures and densities considered and appear unblended with any nearby Sr {\sc ii} lines. In addition the 4-5 transition at 5415.82nm is also of interest as its PEC  is independent of temperature above $0.4$ eV and is strongly correlated with density. The PEC computed for all four of these lines are presented in Figure \ref{fig:5-below} for a range of densities from 10$^6 \leq n_e \leq $ 10$^8$ cm$^{-3}$ and a large temperature range of $T_e$ = 0.1 and 1.0 eV. Clearly these four lines would present excellent density diagnostics for the Y {\sc ii} in the plasma but are not deemed useful for temperature diagnostic work. Given these transitions all originate from the 5th level (4th excited state), and because we know their branching ratios and common upper population, we can determine whether opacities play a key role in altering the calculated Y {\sc ii} densities and abundances from observations. However, the discussed flux calculation at 0.25 eV finds 6.10E+50 $\left( \frac{M_{Y}}{10^{-3} M_\odot}\right)$ ph s$^{-1}$ for the 4071.69 nm, making its observability questionable. Nonetheless, this line exhibits primarily denisity dependent behaviour at a range of temperatures, and could potentially be detected at earlier epochs.

The existence of useful lines for Sr {\sc ii} diagnostics were found to be lacking in comparison. Whilst the 1-2 and 1-3 lines show a density dependence at high temperatures the variation is quite weak. For both Y {\sc ii} and Sr {\sc ii} the majority of lines show a positive temperature dependence, however in most cases the growth over temperature is not a useful metric and most PECs significantly drop off below 0.2 eV, limiting their practicality in diagnosing the highly mixed composition of the KNe spectrum.

\section{Conclusion} \label{sec:conclusions}

New atomic data-sets have been computed for Sr {\sc ii} and Y {\sc ii}. The calculated energy levels and Einstein A-coefficients were found to be in generally good agreement with other theoretical and experimental results in the literature \cite{Biemont11,Palmeri17,Kurucz1995,Hannaford82,Bautista2002}. This gives confidence in the optimised-one-electron orbtials and calculated eigenvectors carried forward to the collision calculations. 

The $R$-matrix method was adopted in both the semi and fully relativistic implementations for the pertinent electron-impact-excitation calculations. The corresponding Maxwellian-averaged-collision strengths were calculated at a large range of temperatures, and will be made publically available in the standard {\sc adf04} format at \cite{openadas_site}. Good convergence in the effective collision strengths was found when tested against smaller models, and in general reasonable agreement was found in electric-dipoles when additionally compared with the approximations of \cite{VanReg1962}. By contrast, forbidden lines saw significant disagreement between the set of $R$-matrix calculations and the approximate formulae optimised for Iron species by \cite{Axelrod1980}. This highlights a potential shortcoming in published NLTE simulations which have employed these approximations. It has been highlighted previously by \cite{Bromley2023} how populations may be misrepresented by such underestimated collision strengths. With the demand for NLTE KNe simulations growing, so too does the need for more robustly calculated collision strengths via the $R$-matrix method.

The newly calculated atomic data-sets were employed in a 1D LTE radiative transfer code {\sc tardis}, where general consistency was found with the previous calculations of G22. In general, it was found that 100\% composition models show sensitivity to individual transitions for Y {\sc ii}, with the Sr {\sc ii} data giving generally consistent results. Additionally, the 4d $\to$ 5p transition array once again shows striking contribution to the spectra in these simulations, supporting the identification from \cite{Watson19}. The emergence of these $\sim 550$ nm features from the 4d$^2$ $\to$ 4d5p transition array of Y {\sc ii} in the 100\% mass composition models is interesting - and reflects the difference in atomic data, where the corresponding spectral lines have oscillator strengths similar to that used by G22, whereas the surrounding lines showed more significant change.

Basic NLTE simulations were carried out via a Collisional-Radiative-Model \citep{bates1962recombination} using the {\sc ColRadPy} \citep{johnson2019colradpy} package. In general, it was found that already highlighted E1 transitions, such as 4d $\to$ 5p transition array of Sr {\sc ii} \citep{Watson19} and the 4d$^2$ $\to$ 4d5p transition array of Y {\sc ii} \citep{SnepWat23} contribute meaningful to the calculated optically thin emission spectra, however forbidden Y {\sc ii} lines were dominant in Kilonova conditions. With these forbidden spectral lines lying in the NIR, they are in principle observable by JWST. Line sensitivities and steady-state populations were calculated and used to predict line luminosities, where we reinforce that Y {\sc ii} is a plausible contributor to the $1.4\mu$m feature, but is unlikely to be the dominant factor.  This presents a potentially interesting and conclusive observation of Y in the KNe should these lines be discernible in future observations. This combined with the above effect on the prevalence of allowed lines suggests a potential need for accurate treatment of forbidden lines in the future. Additionally, the dependence on low-lying forbidden transitions highlights the importance of the replacement of the Axelrod formulae in favour of close-coupling $R$-matrix data as it becomes available.

The data presented here is intended to provide important and accurate representation of these two important $r$-process elements in the modelling of KNe in Binary-Neutron-Star-Mergers.

\section*{Acknowledgements}\label{sec:acknowledgements}

We thank our colleagues at Queen's University Belfast and Auburn University for helpful discussion.  We are grateful for use of the computing resources from the Northern Ireland High Performance Computing (NI-HPC) service funded by EPSRC (EP/T022175). This research made use of \textsc{tardis}, a community-developed software package for spectral synthesis in supernovae \citep{KerzendorfSimTardis, vogl2019,kerzendorf_2023_8244935}. The development of \textsc{tardis} received support from the Google Summer of Code initiative and from ESA’s Summer of Code in Space program. \textsc{tardis} makes extensive use of Astropy and PyNE. Funded/Co-funded by the European Union (ERC, HEAVYMETAL, 101071865). Views and opinions expressed are however those of the author(s) only and do not necessarily reflect those of the European Union or the European Research Council. Neither the European Union nor the granting authority can be held responsible for them. 

\section*{Data Availability}

The effective collision strengths for both Sr {\sc ii} and Y {\sc ii} are recorded in the standard adf04 format on both this articles online supplementary material and at http://open.adas.ac.uk/. Other data underlying this article will be shared on reasonable request to the corresponding author. 


\bibliographystyle{mnras}
\bibliography{bibliography}

\begin{thebibliography}{}
\makeatletter
\relax
\def\mn@urlcharsother{\let\do\@makeother \do\$\do\&\do\#\do\^\do\_\do\%\do\~}
\def\mn@doi{\begingroup\mn@urlcharsother \@ifnextchar [ {\mn@doi@}
  {\mn@doi@[]}}
\def\mn@doi@[#1]#2{\def\@tempa{#1}\ifx\@tempa\@empty \href
  {http://dx.doi.org/#2} {doi:#2}\else \href {http://dx.doi.org/#2} {#1}\fi
  \endgroup}
\def\mn@eprint#1#2{\mn@eprint@#1:#2::\@nil}
\def\mn@eprint@arXiv#1{\href {http://arxiv.org/abs/#1} {{\tt arXiv:#1}}}
\def\mn@eprint@dblp#1{\href {http://dblp.uni-trier.de/rec/bibtex/#1.xml}
  {dblp:#1}}
\def\mn@eprint@#1:#2:#3:#4\@nil{\def\@tempa {#1}\def\@tempb {#2}\def\@tempc
  {#3}\ifx \@tempc \@empty \let \@tempc \@tempb \let \@tempb \@tempa \fi \ifx
  \@tempb \@empty \def\@tempb {arXiv}\fi \@ifundefined
  {mn@eprint@\@tempb}{\@tempb:\@tempc}{\expandafter \expandafter \csname
  mn@eprint@\@tempb\endcsname \expandafter{\@tempc}}}

\bibitem[\protect\citeauthoryear{Alexeeva, Wang, Zhao, Wang, Wu, Wang, Yan  \&
  Shi}{Alexeeva et~al.}{2023}]{Alex2023}
Alexeeva S.,  Wang Y.,  Zhao G.,  Wang F.,  Wu Y.,  Wang J.,  Yan H.,   Shi J.,
   2023, \mn@doi [The Astrophysical Journal] {10.3847/1538-4357/acf5e1}, 957,
  10

\bibitem[\protect\citeauthoryear{Axelrod}{Axelrod}{1980}]{Axelrod1980}
Axelrod T.~S.,  1980, PhD thesis, University of California, Santa Cruz

\bibitem[\protect\citeauthoryear{Badnell}{Badnell}{1986}]{Badnell86}
Badnell N.~R.,  1986, \mn@doi [Journal of Physics B: Atomic and Molecular
  Physics] {10.1088/0022-3700/19/22/023}, 19, 3827

\bibitem[\protect\citeauthoryear{Badnell}{Badnell}{1997}]{Badnell97}
Badnell N.~R.,  1997, \mn@doi [Journal of Physics B: Atomic, Molecular and
  Optical Physics] {10.1088/0953-4075/30/1/005}, 30, 1

\bibitem[\protect\citeauthoryear{Ballance}{Ballance}{2020}]{Ballance}
Ballance C.~P.,  2020, {DARC} {R}-matrix codes,
  \url{http://connorb.freeshell.org}

\bibitem[\protect\citeauthoryear{Bates, Kingston  \& McWhirter}{Bates
  et~al.}{1962}]{bates1962recombination}
Bates D.~R.,  Kingston A.,   McWhirter R.~P.,  1962, Proceedings of the Royal
  Society of London. Series A. Mathematical and Physical Sciences, 267, 297

\bibitem[\protect\citeauthoryear{Bauswein, Goriely  \& Janka}{Bauswein
  et~al.}{2013}]{bauswein_2013}
Bauswein A.,  Goriely S.,   Janka H.-T.,  2013, \mn@doi [The Astrophysical
  Journal] {10.1088/0004-637X/773/1/78}, 773, 78

\bibitem[\protect\citeauthoryear{Bautista, Gull, Ishibashi, Hartman  \&
  Davidson}{Bautista et~al.}{2002}]{Bautista2002}
Bautista M.~A.,  Gull T.~R.,  Ishibashi K.,  Hartman H.,   Davidson K.,  2002,
  \mn@doi [Monthly Notices of the Royal Astronomical Society]
  {10.1046/j.1365-8711.2002.05250.x}, 331, 875

\bibitem[\protect\citeauthoryear{Biémont, Lidberg, Mannervik, Norlin, Royen,
  Schmitt, Shi  \& Tordoir}{Biémont et~al.}{2000}]{Bimont2000}
Biémont E.,  Lidberg J.,  Mannervik S.,  Norlin L.-O.,  Royen P.,  Schmitt A.,
   Shi W.,   Tordoir X.,  2000, \mn@doi [The European Physical Journal D]
  {10.1007/s100530070063}, 11, 355–365

\bibitem[\protect\citeauthoryear{Biémont et~al.,}{Biémont
  et~al.}{2011}]{Biemont11}
Biémont {\'{E}}.,  et~al., 2011, \mn@doi [Monthly Notices of the Royal
  Astronomical Society] {10.1111/j.1365-2966.2011.18637.x}, 414, 3350

\bibitem[\protect\citeauthoryear{Brage, Wahlgren, Johansson, Leckrone  \&
  Proffitt}{Brage et~al.}{1998}]{brage1998theoretical_nist_lines}
Brage T.,  Wahlgren G.~M.,  Johansson S.~G.,  Leckrone D.~S.,   Proffitt C.~R.,
   1998, The Astrophysical Journal, 496, 1051

\bibitem[\protect\citeauthoryear{Bromley, McCann, Loch  \& Ballance}{Bromley
  et~al.}{2023}]{Bromley2023}
Bromley S.~J.,  McCann M.,  Loch S.~D.,   Ballance C.~P.,  2023, \mn@doi [The
  Astrophysical Journal Supplement Series] {10.3847/1538-4365/ace5a1}, 268, 22

\bibitem[\protect\citeauthoryear{Burgess \& Sheorey}{Burgess \&
  Sheorey}{1974}]{Burgess1974}
Burgess A.,  Sheorey V.~B.,  1974, \mn@doi [Journal of Physics B: Atomic and
  Molecular Physics] {10.1088/0022-3700/7/17/026}, 7, 2403

\bibitem[\protect\citeauthoryear{Burgess \& Tully}{Burgess \&
  Tully}{1992}]{Burgess1992analysis}
Burgess A.,  Tully J.,  1992, Astronomy and Astrophysics, Vol. 254, NO. FEB
  (I), P. 436, 1992, 254, 436

\bibitem[\protect\citeauthoryear{Burgess, Chidichimo  \& Tully}{Burgess
  et~al.}{1989}]{Burgess1989}
Burgess A.,  Chidichimo M.~C.,   Tully J.~A.,  1989, \mn@doi [Phys. Rev. A]
  {10.1103/PhysRevA.40.451}, 40, 451

\bibitem[\protect\citeauthoryear{Burke}{Burke}{2011}]{Burke}
Burke P.~G.,  2011, R-Matrix Theory of Atomic Collisions: Application to
  Atomic, Molecular and Optical Processes.
~ Vol. 61, Springer, Verlag Berlin Heidelberg, \url
  {https://www.springer.com/gp/book/9783642159305}

\bibitem[\protect\citeauthoryear{Collins et~al.,}{Collins
  et~al.}{2024}]{collins24}
Collins C.~E.,  et~al., 2024, \mn@doi [Monthly Notices of the Royal
  Astronomical Society] {10.1093/mnras/stae571}, 529, 1333

\bibitem[\protect\citeauthoryear{Dorsch, Latour, Heber, Irrgang, Charpinet  \&
  Jeffery}{Dorsch et~al.}{2020}]{Dorsch2020}
Dorsch M.,  Latour M.,  Heber U.,  Irrgang A.,  Charpinet S.,   Jeffery C.,
  2020, Astronomy \& Astrophysics, 643, A22

\bibitem[\protect\citeauthoryear{Duan, Bari, Wu, Yan  \& Li}{Duan
  et~al.}{2013}]{Duan2013}
Duan B.,  Bari M.~A.,  Wu Z.~Q.,  Yan J.,   Li Y.~M.,  2013, \mn@doi [A\&A]
  {10.1051/0004-6361/201321377}, 555, A144

\bibitem[\protect\citeauthoryear{Dyall, Grant, Johnson, Parpia  \&
  Plummer}{Dyall et~al.}{1989}]{Dyall1989}
Dyall K.,  Grant I.,  Johnson C.,  Parpia F.,   Plummer E.,  1989, \mn@doi
  [Computer Physics Communications] {10.1016/0010-4655(89)90136-7}, 55, 425

\bibitem[\protect\citeauthoryear{Eissner}{Eissner}{1991}]{EISSNER1991}
Eissner W.,  1991, \mn@doi [Le Journal de Physique {IV}] {10.1051/jp4:1991101},
  01, C1

\bibitem[\protect\citeauthoryear{Frebel}{Frebel}{2019}]{FREBEL2019167909}
Frebel A.,  2019, \mn@doi [Annals of Physics]
  {https://doi.org/10.1016/j.aop.2019.167909}, 410, 167909

\bibitem[\protect\citeauthoryear{{Frebel} \& {Ji}}{{Frebel} \&
  {Ji}}{2023}]{Frebel2023}
{Frebel} A.,  {Ji} A.~P.,  2023, \mn@doi [arXiv e-prints]
  {10.48550/arXiv.2302.09188}, \href
  {https://ui.adsabs.harvard.edu/abs/2023arXiv230209188F} {p. arXiv:2302.09188}

\bibitem[\protect\citeauthoryear{{Gillanders}, {Smartt}, {Sim}, {Bauswein}  \&
  {Goriely}}{{Gillanders} et~al.}{2022}]{gillanders2022modelling}
{Gillanders} J.~H.,  {Smartt} S.~J.,  {Sim} S.~A.,  {Bauswein} A.,   {Goriely}
  S.,  2022, \mn@doi [\mnras] {10.1093/mnras/stac1258}, \href
  {https://ui.adsabs.harvard.edu/abs/2022MNRAS.515..631G} {515, 631}

\bibitem[\protect\citeauthoryear{Gillanders, Sim, Smartt, Goriely  \&
  Bauswein}{Gillanders et~al.}{2024}]{gillanders2024modelling}
Gillanders J.~H.,  Sim S.~A.,  Smartt S.~J.,  Goriely S.,   Bauswein A.,  2024,
  \mn@doi [Monthly Notices of the Royal Astronomical Society]
  {10.1093/mnras/stad3688}, p. stad3688

\bibitem[\protect\citeauthoryear{Goriely, Bauswein  \& Janka}{Goriely
  et~al.}{2011}]{goriely_2011}
Goriely S.,  Bauswein A.,   Janka H.-T.,  2011, \mn@doi [The Astrophysical
  Journal Letters] {10.1088/2041-8205/738/2/L32}, 738, L32

\bibitem[\protect\citeauthoryear{Grant, McKenzie, Norrington, Mayers  \&
  Pyper}{Grant et~al.}{1980}]{Grant80}
Grant I.,  McKenzie B.,  Norrington P.,  Mayers D.,   Pyper N.,  1980, \mn@doi
  [Computer Physics Communications] {10.1016/0010-4655(80)90041-7}, 21, 207

\bibitem[\protect\citeauthoryear{{Hannaford}, {Lowe}, {Grevesse}, {Biémont}
  \& {Whaling}}{{Hannaford} et~al.}{1982}]{Hannaford82}
{Hannaford} P.,  {Lowe} R.~M.,  {Grevesse} N.,  {Biémont} {\'{E}}.,
  {Whaling} W.,  1982, \mn@doi [\apj] {10.1086/160384}, \href
  {https://ui.adsabs.harvard.edu/abs/1982ApJ...261..736H} {261, 736}

\bibitem[\protect\citeauthoryear{Hansen, El-Souri, Monaco, Villanova,
  Bonifacio, Caffau  \& Sbordone}{Hansen et~al.}{2018}]{hansen2018ages}
Hansen C.~J.,  El-Souri M.,  Monaco L.,  Villanova S.,  Bonifacio P.,  Caffau
  E.,   Sbordone L.,  2018, The Astrophysical Journal, 855, 83

\bibitem[\protect\citeauthoryear{Hotokezaka, Tanaka, Kato  \&
  Gaigalas}{Hotokezaka et~al.}{2021}]{Hotokezaka2021}
Hotokezaka K.,  Tanaka M.,  Kato D.,   Gaigalas G.,  2021, \mn@doi [Monthly
  Notices of the Royal Astronomical Society] {10.1093/mnras/stab1975}, 506,
  5863

\bibitem[\protect\citeauthoryear{Johnson, Loch  \& Ennis}{Johnson
  et~al.}{2019}]{johnson2019colradpy}
Johnson C.,  Loch S.,   Ennis D.,  2019, Nuclear Materials and Energy, 20,
  100579

\bibitem[\protect\citeauthoryear{{Kerzendorf} \& {Sim}}{{Kerzendorf} \&
  {Sim}}{2014}]{KerzendorfSimTardis}
{Kerzendorf} W.~E.,  {Sim} S.~A.,  2014, \mn@doi [Monthly Notices of the Royal
  Society of Astrophysics] {10.1093/mnras/stu055}, \href
  {https://ui.adsabs.harvard.edu/abs/2014MNRAS.440..387K} {440, 387}

\bibitem[\protect\citeauthoryear{Kerzendorf et~al.,}{Kerzendorf
  et~al.}{2023}]{kerzendorf_2023_8244935}
Kerzendorf W.,  et~al., 2023, tardis-sn/tardis: TARDIS v2023.08.13,
  \mn@doi{10.5281/zenodo.8244935}, \url
  {https://doi.org/10.5281/zenodo.8244935}

\bibitem[\protect\citeauthoryear{Kramida, {Yu.~Ralchenko}, Reader  \& {and NIST
  ASD Team}}{Kramida et~al.}{2023}]{nist}
Kramida A.,  {Yu.~Ralchenko} Reader J.,   {and NIST ASD Team} 2023, {NIST
  Atomic Spectra Database (ver. 5.11), [Online]. Available:
  {\tt{https://physics.nist.gov/asd}} [2024, February 12]. National Institute
  of Standards and Technology, Gaithersburg, MD.}

\bibitem[\protect\citeauthoryear{{Kurucz}}{{Kurucz}}{1995}]{Kurucz1995}
{Kurucz} R.~L.,  1995, in {Adelman} S.~J.,  {Wiese} W.~L.,  eds,  Astronomical
  Society of the Pacific Conference Series Vol. 78, Astrophysical Applications
  of Powerful New Databases. p.~205

\bibitem[\protect\citeauthoryear{Lattimer, Mackie, Ravenhall  \&
  Schramm}{Lattimer et~al.}{1977}]{Latt77}
Lattimer J.,  Mackie F.,  Ravenhall D.,   Schramm 1977, The Astrophysical
  Journal, 213, 225

\bibitem[\protect\citeauthoryear{Madej \& Sankey}{Madej \&
  Sankey}{1990}]{Madej1990}
Madej A.~A.,  Sankey J.~D.,  1990, \mn@doi [Optics Letters]
  {10.1364/ol.15.000634}, 15, 634

\bibitem[\protect\citeauthoryear{Mannervik, Lidberg, Norlin, Royen, Schmitt,
  Shi  \& Tordoir}{Mannervik et~al.}{1999}]{Mannervik1999}
Mannervik S.,  Lidberg J.,  Norlin L.-O.,  Royen P.,  Schmitt A.,  Shi W.,
  Tordoir X.,  1999, \mn@doi [Physical Review Letters]
  {10.1103/physrevlett.83.698}, 83, 698–701

\bibitem[\protect\citeauthoryear{McCann, Bromley, Loch  \& Ballance}{McCann
  et~al.}{2022}]{McCann2022}
McCann M.,  Bromley S.,  Loch S.,   Ballance C.,  2022, Monthly Notices of the
  Royal Astronomical Society, 509, 4723

\bibitem[\protect\citeauthoryear{Moore}{Moore}{1952}]{Moore52}
Moore C.,  1952, Nat.Stand.Ref.Data Ser., Nat.Bur.Stand (U.S.), NSRDS-NBS, II,
  259

\bibitem[\protect\citeauthoryear{Naslim, Jeffery, Behara  \& Hibbert}{Naslim
  et~al.}{2011}]{Naslim2011}
Naslim N.,  Jeffery C.~S.,  Behara N.~T.,   Hibbert A.,  2011, \mn@doi [Monthly
  Notices of the Royal Astronomical Society]
  {10.1111/j.1365-2966.2010.17909.x}, 412, 363

\bibitem[\protect\citeauthoryear{Nilsson, Johansson  \& Kurucz}{Nilsson
  et~al.}{1991}]{Nilsson91}
Nilsson A.~E.,  Johansson S.,   Kurucz R.~L.,  1991, \mn@doi [Physica Scripta]
  {10.1088/0031-8949/44/3/003}, 44, 226

\bibitem[\protect\citeauthoryear{Norrington}{Norrington}{2004}]{Norrington04}
Norrington P.~H.,  2004, DARC, \url {[DEFUNCT]http://www.am.qub.ac.uk/DARC}

\bibitem[\protect\citeauthoryear{Norrington \& Grant}{Norrington \&
  Grant}{1987}]{Norrington87}
Norrington P.~H.,  Grant I.~P.,  1987, \mn@doi [Journal of Physics B: Atomic
  and Molecular Physics] {10.1088/0022-3700/20/18/023}, 20, 4869–4881

\bibitem[\protect\citeauthoryear{OPEN-ADAS}{OPEN-ADAS}{2024}]{openadas_site}
OPEN-ADAS 2024, http://open.adas.ac.uk/, \url {http://open.adas.ac.uk/}

\bibitem[\protect\citeauthoryear{Palmeri, Quinet, Lundberg, Engstr{\"o}m,
  Nilsson  \& Hartman}{Palmeri et~al.}{2017}]{Palmeri17}
Palmeri P.,  Quinet P.,  Lundberg H.,  Engstr{\"o}m L.,  Nilsson H.,   Hartman
  H.,  2017, Monthly Notices of the Royal Astronomical Society, 471, 532

\bibitem[\protect\citeauthoryear{Pian et~al.,}{Pian et~al.}{2017}]{Pian17}
Pian E.,  et~al., 2017, Nature, 551, 67

\bibitem[\protect\citeauthoryear{Pinnington, Berends  \& Lumsden}{Pinnington
  et~al.}{1995}]{Pinnington1995}
Pinnington E.~H.,  Berends R.~W.,   Lumsden M.,  1995, \mn@doi [Journal of
  Physics B: Atomic, Molecular and Optical Physics]
  {10.1088/0953-4075/28/11/009}, 28, 2095–2103

\bibitem[\protect\citeauthoryear{{Pognan}, {Jerkstrand}  \& {Grumer}}{{Pognan}
  et~al.}{2022}]{pognan2022}
{Pognan} Q.,  {Jerkstrand} A.,   {Grumer} J.,  2022, \mn@doi [\mnras]
  {10.1093/mnras/stab3674}, \href
  {https://ui.adsabs.harvard.edu/abs/2022MNRAS.510.3806P} {510, 3806}

\bibitem[\protect\citeauthoryear{Pognan, Grumer, Jerkstrand  \& Wanajo}{Pognan
  et~al.}{2023}]{Pognan2023}
Pognan Q.,  Grumer J.,  Jerkstrand A.,   Wanajo S.,  2023, \mn@doi [Monthly
  Notices of the Royal Astronomical Society] {10.1093/mnras/stad3106}, 526,
  5220

\bibitem[\protect\citeauthoryear{{Reader}, {Corliss}, {Wiese}  \&
  {Martin}}{{Reader} et~al.}{1980}]{reader1980_nist_lines}
{Reader} J.,  {Corliss} C.~H.,  {Wiese} W.~L.,   {Martin} G.~A.,  1980,
  {Wavelengths and transition probabilities for atoms and atomic ions: Part 1.
  Wavelengths, part 2. Transition probabilities}

\bibitem[\protect\citeauthoryear{Sansonetti}{Sansonetti}{2012}]{San12}
Sansonetti J.~E.,  2012, \mn@doi [Journal of Physical and Chemical Reference
  Data] {10.1063/1.3659413}, 41, 013102

\bibitem[\protect\citeauthoryear{Shingles et~al.,}{Shingles
  et~al.}{2019}]{shingles2020}
Shingles L.~J.,  et~al., 2019, \mn@doi [Monthly Notices of the Royal
  Astronomical Society] {10.1093/mnras/stz3412}, 492, 2029

\bibitem[\protect\citeauthoryear{Shingles et~al.,}{Shingles
  et~al.}{2023}]{Shingles2023}
Shingles L.~J.,  et~al., 2023, \mn@doi [The Astrophysical Journal Letters]
  {10.3847/2041-8213/acf29a}, 954, L41

\bibitem[\protect\citeauthoryear{Smartt et~al.,}{Smartt
  et~al.}{2017}]{Smartt2017}
Smartt S.,  et~al., 2017, Nature, 551, 75

\bibitem[\protect\citeauthoryear{Smirnov}{Smirnov}{2001}]{Smirnov2001}
Smirnov Y.~M.,  2001, \mn@doi [High Temperature] {10.1023/A:1013126602902}, 39,
  815

\bibitem[\protect\citeauthoryear{Sneppen \& Watson}{Sneppen \&
  Watson}{2023}]{SnepWat23}
Sneppen A.,  Watson D.,  2023, \mn@doi [A\&A] {10.1051/0004-6361/202346421},
  675, A194

\bibitem[\protect\citeauthoryear{Storm \& Bergemann}{Storm \&
  Bergemann}{2023}]{StormBerg2023}
Storm N.,  Bergemann M.,  2023, \mn@doi [Monthly Notices of the Royal
  Astronomical Society] {10.1093/mnras/stad2488}, 525, 3718

\bibitem[\protect\citeauthoryear{Storm et~al.,}{Storm et~al.}{2024}]{Storm2024}
Storm N.,  et~al., 2024, \mn@doi [A\&A Forthcoming article]
  {https://doi.org/10.1051/0004-6361/202348971}

\bibitem[\protect\citeauthoryear{Summers et~al.,}{Summers
  et~al.}{2006}]{Summers2006}
Summers H.~P.,  et~al., 2006, \mn@doi [Plasma Physics and Controlled Fusion]
  {10.1088/0741-3335/48/2/007}, 48, 263

\bibitem[\protect\citeauthoryear{Tanaka, Kato, Gaigalas  \& Kawaguchi}{Tanaka
  et~al.}{2020}]{Tanaka20}
Tanaka M.,  Kato D.,  Gaigalas G.,   Kawaguchi K.,  2020, \mn@doi [Monthly
  Notices of the Royal Astronomical Society] {10.1093/mnras/staa1576}, 496,
  1369

\bibitem[\protect\citeauthoryear{{Vogl}, {Sim}, {Noebauer}, {Kerzendorf}  \&
  {Hillebrandt}}{{Vogl} et~al.}{2019}]{vogl2019}
{Vogl} C.,  {Sim} S.~A.,  {Noebauer} U.~M.,  {Kerzendorf} W.~E.,
  {Hillebrandt} W.,  2019, \mn@doi [\aap] {10.1051/0004-6361/201833701}, \href
  {https://ui.adsabs.harvard.edu/abs/2019A&A...621A..29V} {621, A29}

\bibitem[\protect\citeauthoryear{Watson et~al.,}{Watson
  et~al.}{2019}]{Watson19}
Watson D.,  et~al., 2019, \mn@doi [Nature] {10.1038/s41586-019-1676-3}, 574,
  497–500

\bibitem[\protect\citeauthoryear{Wiese \& Martin}{Wiese \&
  Martin}{1980}]{WieMar80}
Wiese W.~L.,  Martin G.~A.,  1980, in , Wavelengths and Transition
  Probabilities for Atoms and Atomic Ions, Nat. Stand. Ref. Data Ser. NSRDS-68.
Nat. Bur. Stand., U.S., \mn@doi{10.6028/NBS.NSRDS.68}

\bibitem[\protect\citeauthoryear{{van Regemorter}}{{van
  Regemorter}}{1962}]{VanReg1962}
{van Regemorter} H.,  1962, \mn@doi [\apj] {10.1086/147445}, \href
  {https://ui.adsabs.harvard.edu/abs/1962ApJ...136..906V} {136, 906}

\makeatother
\end{thebibliography}






\bsp	
\label{lastpage}
\end{document}